\documentclass{ieeeaccess}
\usepackage{cite}
\usepackage{amsmath,amssymb,amsfonts}
\usepackage{algorithmic}
\usepackage{graphicx}
\usepackage{textcomp}
\usepackage{bm}
\usepackage{multirow}
\usepackage{booktabs}
\usepackage{colortbl}
\definecolor{tabhl}{RGB}{217,232,252}
\usepackage{hyperref}

\newcommand{\fitcolumn}[1]{\resizebox{\ifdim\width>\columnwidth \columnwidth\else\width\fi}{!}{#1}}

\makeatletter
\AtBeginDocument{\DeclareMathVersion{bold}
\SetSymbolFont{operators}{bold}{T1}{times}{b}{n}
\SetSymbolFont{NewLetters}{bold}{T1}{times}{b}{it}
\SetMathAlphabet{\mathrm}{bold}{T1}{times}{b}{n}
\SetMathAlphabet{\mathit}{bold}{T1}{times}{b}{it}
\SetMathAlphabet{\mathbf}{bold}{T1}{times}{b}{n}
\SetMathAlphabet{\mathtt}{bold}{OT1}{pcr}{b}{n}
\SetSymbolFont{symbols}{bold}{OMS}{cmsy}{b}{n}
\renewcommand\boldmath{\@nomath\boldmath\mathversion{bold}}}
\makeatother

\def\BibTeX{{\rm B\kern-.05em{\sc i\kern-.025em b}\kern-.08em
    T\kern-.1667em\lower.7ex\hbox{E}\kern-.125emX}}

\def\headerlogo{\raisebox{-2pt}{\includegraphics[width=7.61pc,height=18pt,keepaspectratio=false]{logo.png}}}
\def\headerlogoall{\raisebox{-2pt}{\includegraphics[width=7.61pc,height=18pt,keepaspectratio=false]{notaglinelogo.png}}}

\begin{document}
\history{Date of publication xxxx 00, 0000, date of current version xxxx 00, 0000.}
\doi{10.1109/ACCESS.2024.0429000}

\title{ResMamba: A Dual-Path Convolution-Mamba Model for Music Genre Classification}

\author{
\uppercase{Yingyu Zhang}\authorrefmark{1}\authorrefmark{3}, 
\uppercase{Pu Qu}\authorrefmark{2}\authorrefmark{3}
}

\address[1]{School of Artificial Intelligence, Hainan Normal University, Haikou 571127 China (e-mail: 240120@hainnu.edu.cn)}
\address[2]{School of Teacher Education, Hainan Normal University, Haikou 571127 China  (e-mail: qupu1983@163.com)}

\markboth
{Zhang \headeretal: ResMamba: A Dual-Path Convolution-Mamba Model for Music Genre Classification}
{Zhang \headeretal: ResMamba: A Dual-Path Convolution-Mamba Model for Music Genre Classification}

\corresp{Corresponding author: Yingyu Zhang (e-mail: 240120@hainnu.edu.cn).}

\tfootnote{This work was supported in part by the Hainan Natural Science Foundation of China under Grant 723QN237.\\\authorrefmark{3}Co-corresponding author: Pu Qu (e-mail: qupu1983@163.com).}

\begin{abstract}
Music genre classification remains a fundamental yet challenging task in music information retrieval, requiring models that capture local time--frequency patterns and long-range temporal dependencies. To address this, we propose ResMamba, a hybrid architecture that augments a ResNet-style four-stage backbone with Mamba blocks. Building upon the selective structured state space sequence model, ResMamba first adapts the selective SSM into a 4D-compatible residual Mamba block that captures global dependencies at linear complexity. We then design the ConvMamba dual-path block, which combines this adapted Mamba branch with a parallel convolutional branch through a learned $1\times1$ fusion, jointly modeling global sequential context and local spectro-temporal patterns. Furthermore, we adopt a multi-scale fusion head that aggregates hierarchical representations across all four stages. Experiments on the GTZAN dataset demonstrate that our full ResMamba model achieves a 5-fold cross-validation accuracy of $88.99\% \pm 1.58\%$, surpassing the strongest published method in our comparison (88.64\%). All models are trained from scratch without any pretraining, external data, or data augmentation. Beyond GTZAN, ResMamba also generalizes to the larger FMA benchmarks under the same training protocol, attaining $64.52\% \pm 1.88\%$ on FMA-Small under 10-fold cross-validation --- the best published from-scratch result --- and $65.75\%$ on FMA-Medium under the official artist-level split, surpassing the widely cited SVM baseline of Defferrard \textit{et al.} (63\%). 
\end{abstract}


\titlepgskip=-21pt
\maketitle

\section{Introduction}
\label{sec:introduction}

Music genre classification remains a cornerstone task in Music Information Retrieval (MIR). The ability to automatically categorize musical pieces into predefined genres---such as classical, jazz, rock, or pop---underpins a wide range of applications, including music streaming services, automated playlist generation, content-based recommendation systems, and large-scale archival management. Since the introduction of the GTZAN dataset~\cite{gtzan} and systematic feature-engineering approaches by Tzanetakis and Cook, the field has undergone a significant paradigm shift: evolving from hand-crafted acoustic features with shallow classifiers to deep learning methods capable of learning representations directly from raw or preprocessed audio.

Convolutional Neural Networks (CNNs) have long served as the dominant paradigm for spectrogram-based music classification. Architectures such as ResNet~\cite{resnet}, VGG~\cite{vgg}, and CRNNs~\cite{choi2016} treat mel-spectrograms as two-dimensional images, applying successive local convolutions to extract time--frequency patterns. Despite their effectiveness, the intrinsic inductive bias of CNNs---specifically, the limited receptive field of convolutional kernels---imposes an architectural bottleneck on modeling long-range dependencies across musical sections. Genre-defining characteristics, such as verse--chorus transitions, instrumentation variations, and rhythmic modulations, often unfold over tens of seconds, necessitating the ability to correlate temporally distant events along the sequence axis.

Among CNN architectures, ResNet~\cite{resnet} stands as one of the most influential designs in modern deep learning. By introducing residual connections with identity mappings, ResNet mitigates the degradation problem that hindered earlier deep networks, enabling stable training of very deep architectures. It has been successfully adapted to diverse modalities such as speech recognition, environmental sound classification, and music information retrieval, where its hierarchical feature extraction paradigm---characterized by four stages with progressively increasing channel dimensions and decreasing spatial resolutions---provides a natural multi-scale representation that aligns well with the compositional structure of audio signals.

The recent emergence of State Space Models (SSMs), particularly the Structured State Space Sequence Model (S4)~\cite{s4} and its selective variant, Mamba (S6)~\cite{mamba}, has opened new avenues for sequence modeling with linear complexity. Mamba introduces a data-dependent selection mechanism that enables the model to selectively retain relevant tokens while filtering out noise, achieving performance competitive with Transformers on language and vision tasks while maintaining $O(N)$ inference complexity. Subsequent works, including Vision Mamba~\cite{visionmamba} and VMamba~\cite{vmamba}, have further demonstrated that the Mamba framework can be effectively adapted to visual recognition through multi-directional scanning strategies for image patches.

In this paper, we propose \textbf{ResMamba}, a hybrid architecture that integrates the selective SSM block of Mamba into a ResNet-style four-stage hierarchical backbone for music genre classification. Our key contributions are threefold:

\begin{enumerate}
    \item We design a \textbf{4D-compatible Mamba wrapper} that encapsulates the selective SSM within a 4D residual module interface, \texttt{MambaWrapper}, which accepts and returns feature maps of shape $(B, C, H, W)$. This module enables global context aggregation with linear time complexity while maintaining compatibility with hierarchical CNN backbones.
    
    \item We propose a \textbf{ConvMamba dual-path block} comprising two parallel branches: a Mamba branch leveraging the aforementioned 4D-adapted module for global sequence modeling, and a convolutional branch employing two stacked $3\times3$ convolutions for local spectro-temporal feature extraction. The outputs from both branches are concatenated and fused via a learnable $1\times1$ convolution, effectively preserving complementary global and local representations.
    
    \item We adopt a \textbf{multi-scale fusion head} that aggregates hierarchical features from all four stages prior to classification; ablation studies confirm that this aggregation yields complementary gains over the proposed ConvMamba blocks alone.
\end{enumerate}

Extensive experiments on the GTZAN dataset validate the effectiveness of our approach. Notably, the ConvMamba variant with multi-scale fusion achieves a 5-fold cross-validation accuracy of $88.99\% \pm 1.58\%$, trained from scratch without any pretraining, external data, or data augmentation. Beyond GTZAN, ResMamba also generalizes to the larger Free Music Archive (FMA) benchmarks under the same protocol, attaining $64.52\% \pm 1.88\%$ on FMA-Small under 10-fold cross-validation and $65.75\%$ on FMA-Medium under the official artist-level split.

The remainder of this paper is organized as follows. Section~\ref{sec:related} reviews related work in music genre classification and state space models. Section~\ref{sec:method} details the proposed ResMamba architecture. Section~\ref{sec:experiments} describes the experimental setup and presents a comprehensive analysis of the results. Finally, Section~\ref{sec:conclusion} concludes the paper and outlines directions for future work.

\section{Related Work}
\label{sec:related}

\subsection{Music Genre Classification}

The GTZAN dataset, introduced by Tzanetakis and Cook~\cite{gtzan}, has served as one of the most widely used benchmarks for music genre classification for over two decades. Early approaches predominantly relied on hand-engineered acoustic features---including mel-frequency cepstral coefficients (MFCCs), spectral centroid, zero-crossing rate, and rhythmic descriptors---coupled with classical classifiers such as support vector machines (SVMs) or Gaussian mixture models (GMMs). More recently, Muniz and Souza Filho~\cite{muniz2023} revisited hand-crafted feature pipelines with a statistically grounded wrapper-based feature selection procedure, deriving a compact feature set that transfers across GTZAN, FMA, and other music datasets. In a complementary statistical direction, Shen and Xiao~\cite{shen2024fda} formulated genre classification as a functional data analysis problem, modeling the temporal trajectories of acoustic features with basis-function expansions and thereby offering a principled non-deep alternative on standard benchmarks.

The advent of deep learning catalyzed significant progress in this domain. Choi \emph{et al.}~\cite{hcnn} pioneered the application of CNNs directly to mel-spectrograms, reframing audio classification as an image recognition task. Kim \emph{et al.}~\cite{kim2020nc} subsequently demonstrated through extensive multi-task transfer-learning experiments that the choice and number of learning source tasks substantially influence the quality of deep music representations. This paradigm was subsequently advanced through deeper architectures borrowed from computer vision, including VGG~\cite{vgg}, ResNet~\cite{resnet}, DenseNet~\cite{densenet}, MobileNetV2~\cite{mobilenetv2}, MobileNetV3~\cite{mobilenetv3}, and ShuffleNetV2~\cite{shufflenetv2}. More recently, attention mechanisms, particularly self-attention within Transformer encoders~\cite{wu2025vit,chen2024cnnte}, have been investigated for their capacity to capture long-range dependencies in music; however, the quadratic computational complexity of standard attention remains a bottleneck for scalability on extended audio sequences.

Recurrent neural networks (RNNs), notably long short-term memory (LSTM) networks, have also been employed to model the temporal evolution of audio features. In particular, convolutional recurrent neural networks~\cite{choi2016} pair a convolutional front-end with a recurrent sequence model to capture both spectro-temporal patterns and long-term temporal context. Transfer learning has further amplified the utility of such deep representations: Choi \emph{et al.}~\cite{crnn_music} showed that models trained on large-scale audio tagging data transfer effectively to classification and regression tasks. Nevertheless, LSTMs are inherently limited by vanishing gradients over prolonged sequences and sequential processing constraints, which hinder training efficiency on modern parallel hardware.

Complementary to fully supervised audio-only modeling, recent studies explore label-efficient and knowledge-augmented settings. Shen and Xiao~\cite{shen2026} adapt frozen music foundation models to zero-to-few-shot genre classification via prototype calibration and embedding-space prompt learning, while Ding \emph{et al.}~\cite{ding2024} embed genre-centric knowledge graphs into audio representations, achieving 68.07\% accuracy on FMA-Medium. Although these methods achieve competitive performance, they depend on external semantic knowledge, pretrained speech-language backbones, or auxiliary metadata, rather than compact end-to-end audio architectures.

\subsection{State Space Models and Mamba}

State space models (SSMs) offer a principled framework for modeling continuous-time dynamics and have recently re-emerged as efficient alternatives to attention mechanisms for long-sequence modeling. The S4 model~\cite{s4} introduced structured diagonal parameterization of the state matrix, achieving $O(N \log N)$ complexity in the convolutional mode, with the state matrix initialized via HiPPO. Subsequent variants, including S4D~\cite{s4d} and DSS~\cite{dss}, further streamlined this parameterization while preserving strong performance on the Long Range Arena benchmark.

A pivotal advancement came with Mamba~\cite{mamba}, which incorporated an input-dependent selection mechanism into the SSM framework. This selective structured state space model (S6) endows the network with content-aware reasoning capabilities: the discretized state transition matrices $\mathbf{A}$ and $\mathbf{B}$ become functions of the input token, enabling dynamic propagation or filtering of information along the sequence. Consequently, Mamba achieves Transformer-comparable quality on language modeling tasks while maintaining linear scaling with respect to sequence length.

In the vision domain, Vision Mamba (VIM)~\cite{visionmamba} and VMamba~\cite{vmamba} adapted the Mamba block for image recognition. VIM employs bidirectional scanning over flattened image patches to capture global visual context, whereas VMamba introduces a cross-scan module that traverses 2D feature maps along four directions---horizontal, vertical, and two diagonals---to enrich spatial representation. These efforts demonstrated that Mamba-based architectures can match ViT and Swin Transformer variants on ImageNet classification while offering substantial memory efficiency gains.

Hybrid convolution--SSM designs have also been explored beyond natural-image recognition. MedMamba~\cite{medmamba} couples convolutional layers with the selective SSM for medical image classification, and a homonymous \emph{ConvMamba}~\cite{convmamba_hsi} combines Mamba with CNNs for hyperspectral image classification. Although these models share the philosophy of complementing the SSM with convolutional inductive biases, our ConvMamba block is architecturally distinct: it embeds a 4D-adapted Mamba branch and a parallel two-layer convolutional branch within a residual block, fuses the two branches via channel concatenation followed by a learned $1\times1$ convolution. This design enables more effective capture of time-frequency patterns from the anisotropic 2D Mel spectrogram.

Building upon these foundations, our work adapts Mamba specifically for music spectrograms, where the temporal axis encodes musical structure and the frequency axis captures pitch and timbral information. Unlike natural images, where spatial translation invariance is paramount, mel-spectrograms exhibit anisotropic semantics: the horizontal (time) dimension conveys long-range musical narrative, while the vertical (frequency) dimension reflects localized harmonic structure. ResMamba is explicitly designed to respect this asymmetry through its dual-path convolution-SSM architecture.

\section{Proposed Method: ResMamba}
\label{sec:method}

\subsection{Overall Architecture}

\begin{figure}[t]
\centering
\includegraphics[width=0.9\linewidth]{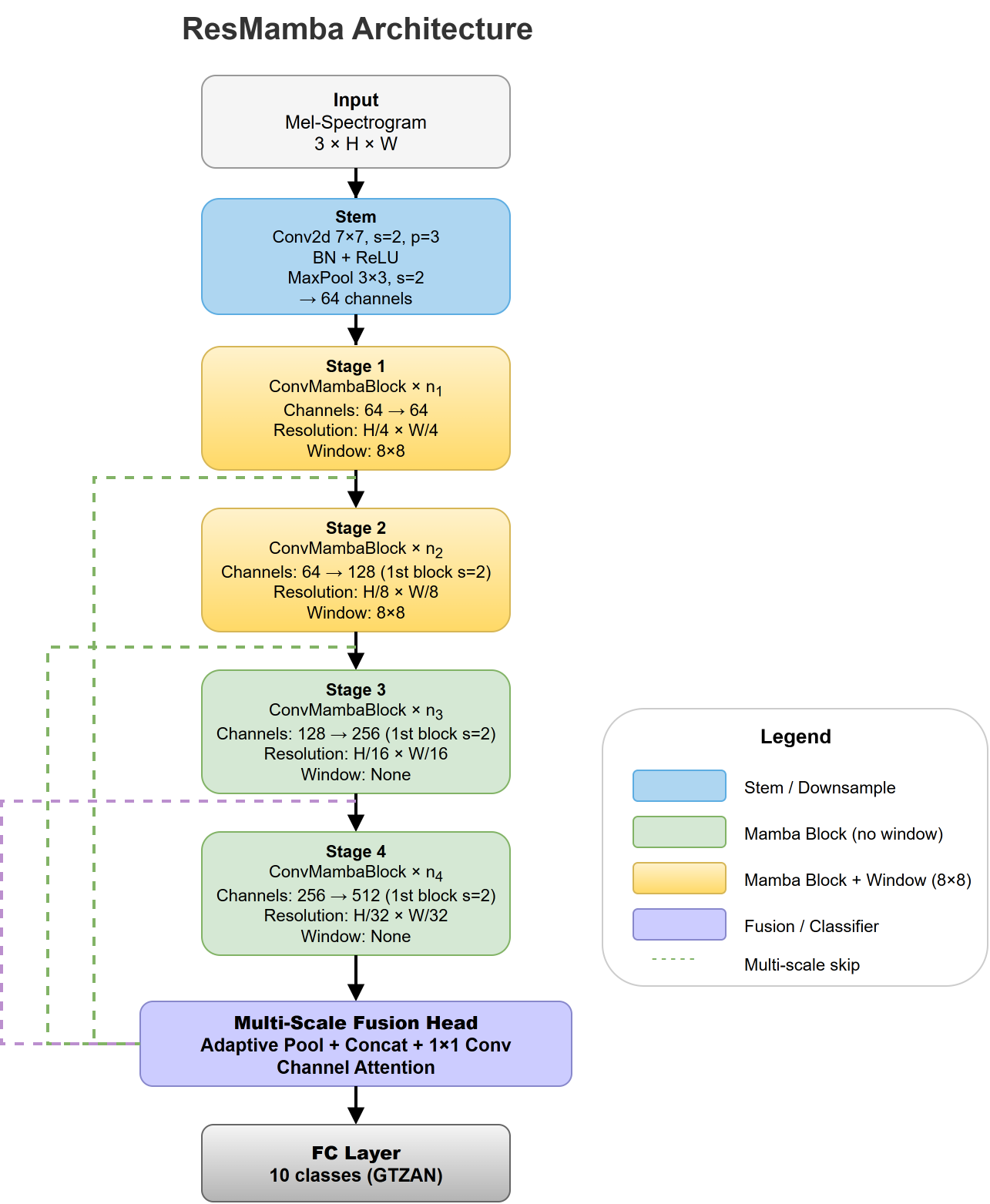}
\caption{Overall architecture of ResMamba. The backbone adopts a four-stage hierarchy, beginning with a stem that projects the 3-channel mel-spectrogram input to 64 feature maps via a $7\times7$ convolution followed by a $3\times3$ max-pooling layer, yielding a feature map of size $64 \times H/4 \times W/4$. In the high-resolution Stages~1 and~2, whose flattened feature maps would form very long sequences, the feature maps are partitioned into non-overlapping soft windows of size $8\times8$ to keep the per-scan sequence length tractable, whereas the low-resolution Stages~3 and~4, whose feature maps are already downsampled to $H/16 \times W/16$ and $H/32 \times W/32$, allow full-sequence SSM to capture global context across the entire feature map with modest computational overhead. A multi-scale fusion head aggregates features from all stages prior to classification.}

\label{fig:architecture}
\end{figure}

ResMamba employs a hierarchical ResNet-style backbone comprising four stages, as illustrated in Fig.~\ref{fig:architecture}.    Given an input mel-spectrogram $\mathbf{X} \in \mathbb{R}^{3 \times H \times W}$, the stem first projects it to a 64-channel feature map via a $7\times7$ convolution with stride 2 and padding 3, followed by batch normalization and ReLU activation:
\begin{equation}
    \mathbf{F}_0 = \text{ReLU}\big(\text{BN}\big(\text{Conv2d}_{7\times7,\, s=2,\, p=3}(\mathbf{X})\big)\big),
    \label{eq:stem}
\end{equation}
and subsequently applies a $3\times3$ max-pooling layer with stride 2, yielding a feature map of size  $64 \times H/4 \times W/4$.

This feature map is subsequently processed through four sequential stages. Each stage $k \in \{1,2,3,4\}$ comprises $n_k$ residual blocks and produces $C_k$ output channels, with $(C_1, C_2, C_3, C_4) = (64, 128, 256, 512)$; spatial downsampling is applied at the beginning of stages~2--4. In principle, any building block that accepts and returns 4D feature maps can be integrated into this architecture, including standard ResNet BasicBlock or Bottleneck modules. While each stage permits flexible instantiation of its core component, this work specifically employs two proposed variants: the Mamba block (\texttt{AdaptMambaBlock}), which substitutes the two stacked convolutions in the native ResNet BasicBlock with a \texttt{MambaWrapper} followed by BN, and the ConvMamba dual-path module (\texttt{ConvMambaBlock}), which integrates this Mamba branch (\texttt{MambaWrapper} + BN) with a parallel convolutional branch.

\subsection{MambaWrapper: Wrapping Selective SSM for 2D Features}
\label{sec:mamba_block}

We adopt the unidirectional MambaVision mixer~\cite{mambavision} and encapsulate it within a 4D residual module interface, \texttt{MambaWrapper}, which accepts and returns feature maps of shape $(B, C, H, W)$. The native Mamba block operates on 3D sequences $\mathbf{x} \in \mathbb{R}^{B \times N \times D}$, where $B$ is the batch size, $N$ the sequence length, and $D$ the token embedding dimension. To bridge this dimensional mismatch with 2D spectrogram features $\mathbf{F} \in \mathbb{R}^{B \times C \times H \times W}$, \texttt{MambaWrapper} performs a structure-preserving round-trip transformation. Specifically, it first applies a $1\times1$ convolution (\texttt{proj\_in}) to project the channel dimension from $C$ to $D$ while maintaining the 4D spatial layout:

\begin{equation}
\mathbf{F}_p = \text{Conv2d}_{1\times1}(\mathbf{F}) \in \mathbb{R}^{B \times D \times H \times W}
\end{equation}

The projected 4D tensor $\mathbf{F}_p$ is then flattened along the spatial axes and transposed to form a 3D token sequence with sequence length $N = H \cdot W$:

\begin{equation}
\mathbf{u} = \text{Flatten}(\mathbf{F}_p) \in \mathbb{R}^{B \times N \times D}
\end{equation}

The resulting 3D tensor $\mathbf{u}$ is then fed into the native selective SSM (MambaVisionMixer). Internally, the SSM operates as follows. The input $\mathbf{u}$ is projected to an inner dimension $2d_{\text{inner}}$ through a linear projection and split into two branches $\mathbf{x}$ and $\mathbf{z}$, each of width $d_{\text{inner}}$:

\begin{equation}
\mathbf{x}, \mathbf{z} = \text{Split}(\text{Proj}_{\text{in}}(\mathbf{u}))
\end{equation}

Each branch passes through a depthwise 1D convolution ($d_{\text{conv}}=3$) followed by SiLU activation to capture local sequential context:

\begin{equation}
\tilde{\mathbf{x}} = \text{SiLU}(\text{DWConv1d}(\mathbf{x})), \quad
\tilde{\mathbf{z}} = \text{SiLU}(\text{DWConv1d}(\mathbf{z}))
\end{equation}

The selective SSM parameters are computed from $\tilde{\mathbf{x}}$:

\begin{align}
\mathbf{B} &= \text{Linear}_B(\tilde{\mathbf{x}}) \in \mathbb{R}^{L \times d_{\text{state}}} \\
\mathbf{C} &= \text{Linear}_C(\tilde{\mathbf{x}}) \in \mathbb{R}^{L \times d_{\text{state}}} \\
\boldsymbol{\Delta} &= \text{Softplus}(\text{Linear}_{\Delta}(\tilde{\mathbf{x}}) + \mathbf{b}_{\Delta})
\label{eq:ssm_params}
\end{align}

where $d_{\text{state}} = 8$ is the state dimension. The continuous-time parameters $\mathbf{A}, \mathbf{B}, \mathbf{C}$ are discretized with step size $\boldsymbol{\Delta}$ using the first-order approximation:

\begin{equation}
\overline{\mathbf{A}} = \exp(\boldsymbol{\Delta} \mathbf{A}), \quad
\overline{\mathbf{B}} = \boldsymbol{\Delta} \mathbf{B}
\label{eq:discretize}
\end{equation}

where $\mathbf{A} \in \mathbb{R}^{d_{\text{inner}} \times d_{\text{state}}}$ has a diagonal structure with negative real eigenvalues to guarantee stability. The SSM computes the output via the recurrent formulation:

\begin{equation}
\mathbf{h}_{t} = \overline{\mathbf{A}}_t \mathbf{h}_{t-1} + \overline{\mathbf{B}}_t \tilde{\mathbf{x}}_t, \quad
\mathbf{y}_t = \mathbf{C}_t \mathbf{h}_t + \mathbf{D} \odot \tilde{\mathbf{x}}_t
\label{eq:ssm_recurrence}
\end{equation}

where the subscript $t$ indicates that $\overline{\mathbf{A}}_t$, $\overline{\mathbf{B}}_t$, and $\mathbf{C}_t$ are input-dependent and vary across positions. This selective mechanism is the core innovation of S6 over prior SSM architectures, as it allows the model to dynamically decide which information to retain or discard along the sequence.

The final output is formed by concatenating the SSM output with the $\tilde{\mathbf{z}}$ branch and projecting the combined representation back to the embedding dimension:

\begin{equation}
\mathbf{o} = \text{Proj}_{\text{out}}\left(\text{Concat}\left(\mathbf{y}, \tilde{\mathbf{z}}\right)\right)
\label{eq:gate}
\end{equation}

where $\mathbf{D}$ is a learnable per-channel skip-connection parameter, and $\text{Proj}_{\text{out}}$ projects the concatenated $2d_{\text{inner}}$-dimensional representation back to dimension $D$. Finally, the \texttt{MambaWrapper} recovers the 4D spatial layout by transposing and reshaping $\mathbf{o}$ from $(B, N, D)$ to $(B, D, H, W)$, completing the dimension-preserving round-trip transformation.

To balance global context modeling with computational efficiency, we employ a soft window-partitioning strategy. In the high-resolution stages (Stages 1 and 2), flattening the entire feature map into a single sequence would produce very long token streams ($H\cdot W/16$ and $H\cdot W/64$, i.e., $32\times323=10{,}336$ and $16\times161=2{,}576$ tokens for the $128\times1292$ mel-spectrogram input); we therefore partition the feature map into non-overlapping windows of size $8\times8$ and execute the SSM independently within each window. This reduces the per-scan sequence length to 64 tokens and, for the $128\times1292$ mel-spectrogram input, confines each window to roughly 0.7--1.5\,s of audio, so that short rhythmic and phrasal structures are captured locally at low cost. In the low-resolution, high-channel stages (Stages 3 and 4), whose feature maps contain only $H\cdot W/256$ and $H\cdot W/1024$ tokens (640 and 160 for the $128\times1292$ input), we instead apply full-sequence SSM ($\text{window\_size}=0$) to aggregate global context over the entire feature map with modest computational overhead. Unlike Swin Transformer's hard shifted-window partitioning, our soft windows are non-overlapping and are partitioned only once, with the window grid kept fixed throughout the network: local context inside each window is enriched by the depthwise convolution preceding the SSM, while cross-window information is re-integrated through the parallel convolutional branch of the ConvMamba block and the multi-scale aggregation across stages. No shifted-window re-partitioning is required.

\subsection{AdaptMambaBlock: Single-Path Mamba Residual Block}
\label{sec:mamba_block_single}

\begin{figure}[t]
\centering
\includegraphics[width=0.9\linewidth]{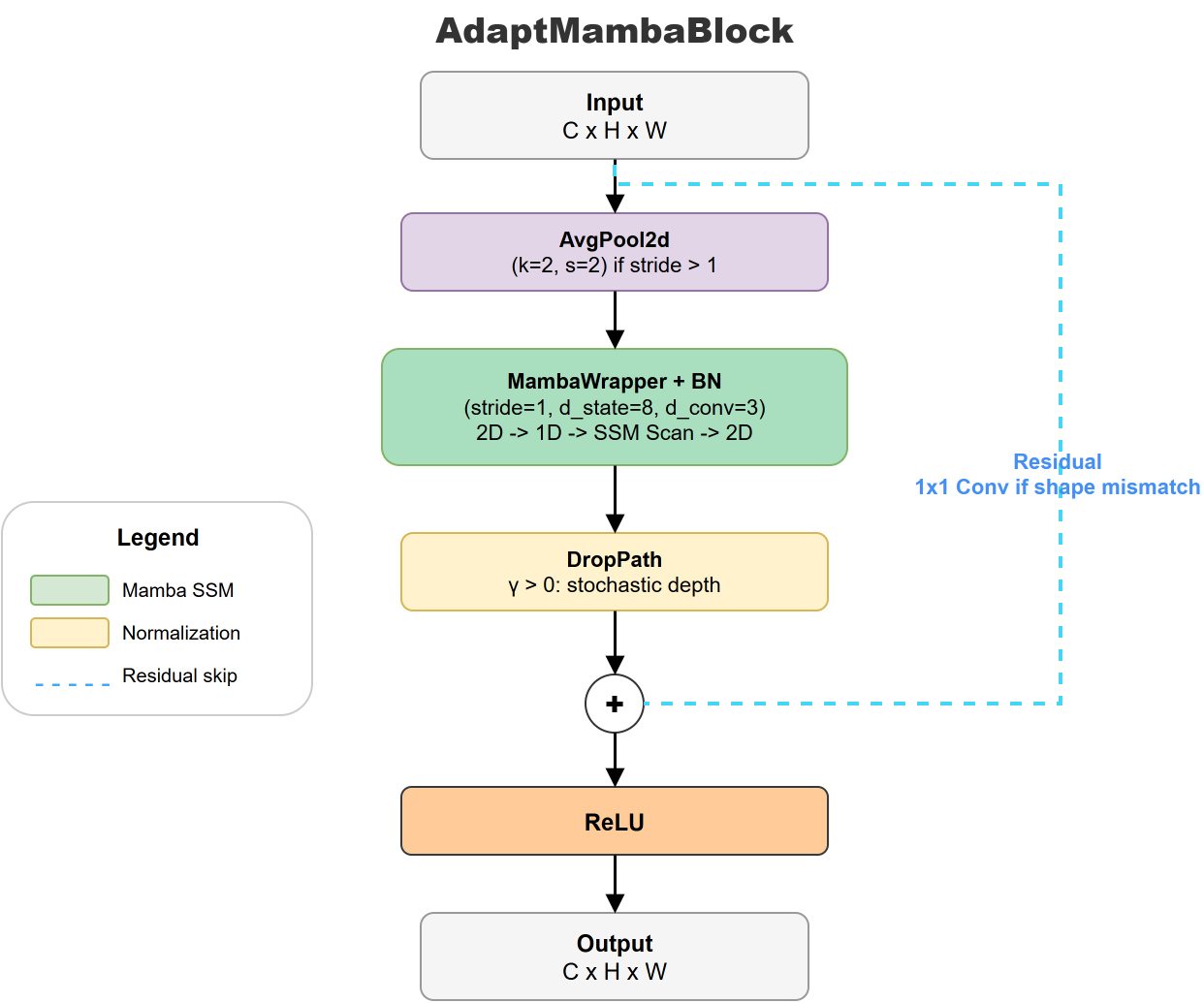}
\caption{Architecture of the \texttt{AdaptMambaBlock}. The input is optionally downsampled via AvgPool2d, then processed by a \texttt{MambaWrapper} (stride=1) for global sequential modeling, followed by BatchNorm2d and DropPath regularization. A residual connection adds the identity (or $1\times1$-convolved) input, and ReLU produces the final output.}
\label{fig:mamba_block}
\end{figure}

As the most elementary instantiation of the Mamba SSM in the ResNet-style framework, \texttt{AdaptMambaBlock} serves as a drop-in replacement for the standard ResNet basic blocks (e.g., BasicBlock or Bottleneck) that are stacked to form each stage, replacing their convolutional pathway with Mamba-based state-space sequence modeling. 
Configured with an expansion factor of $1$ (aligning with the BasicBlock design), it establishes a lightweight, single-path architecture that serves as a baseline for assessing the specific contributions of selective state-space mechanisms to music genre classification. The detailed architecture is illustrated in Fig.~\ref{fig:mamba_block} and formally defined below.

Given an input feature map $\mathbf{x} \in \mathbb{R}^{C \times H \times W}$, optional spatial downsampling is applied when inter-stage resolution reduction is required ($\text{stride}>1$):
\begin{equation}
    \mathbf{x}_s = \operatorname{AvgPool2d}_{2\times2}(\mathbf{x}),
    \label{eq:mamba_ds}
\end{equation}
where $\mathbf{x}_s = \mathbf{x}$ when $\text{stride}=1$. The resulting representation is then processed by a \texttt{MambaWrapper} module. To preserve structural integrity during the 2D$\to$1D$\to$2D round-trip transformation, this module strictly operates with $\text{stride}=1$, ensuring proper alignment between 2D spatial dimensions and the 1D token sequence:
\begin{equation}
    \mathbf{x}_m = \operatorname{BN}\bigl(\operatorname{MambaWrapper}_{\text{stride}=1}(\mathbf{x}_s)\bigr).
    \label{eq:mamba_main}
\end{equation}

Stochastic depth regularization (DropPath) is subsequently applied to enhance training stability and generalization:
\begin{equation}
    \tilde{\mathbf{x}}_m = \operatorname{DropPath}_\gamma(\mathbf{x}_m),
    \label{eq:mamba_droppath}
\end{equation}
where the drop probability $\gamma$ increases linearly with block depth. A residual shortcut provides the identity path; when spatial dimensions or channel counts mismatch ($\text{stride}>1$), a $1\times1$ convolution with stride 2 followed by BatchNorm is employed to align the skip connection:
\begin{equation}
    \mathbf{x}_{\text{skip}} = 
    \begin{cases}
        \operatorname{BN}\!\bigl(\operatorname{Conv2d}_{1\times1,\,s=2}(\mathbf{x})\bigr), & \text{if } \text{stride}>1, \\[4pt]
        \mathbf{x}, & \text{otherwise}.
    \end{cases}
    \label{eq:mamba_skip}
\end{equation}

Finally, the output is obtained through residual addition followed by activation:
\begin{equation}
    \mathbf{x}_{\text{out}} = \operatorname{ReLU}\bigl(\tilde{\mathbf{x}}_m + \mathbf{x}_{\text{skip}}\bigr).
    \label{eq:mamba_output}
\end{equation}

In contrast to the dual-path ConvMamba block (Section~\ref{sec:convmamba}), \texttt{AdaptMambaBlock} relies solely on the Mamba branch, omitting any parallel convolutional path. This yields a more streamlined architecture with reduced parameter count and computational overhead. Furthermore, the $\text{expansion}=1$ configuration maintains a constant channel dimension throughout the feed-forward transformation, preserving the parameter efficiency characteristic of the original ResNet BasicBlock. Such a design facilitates direct ablation comparisons between pure Mamba-based and hybrid ConvMamba residual architectures.

\subsection{ConvMambaBlock: Dual-Path Convolution-Mamba Residual Block}
\label{sec:convmamba}

\begin{figure}[t]
    \centering
    \includegraphics[width=0.9\linewidth]{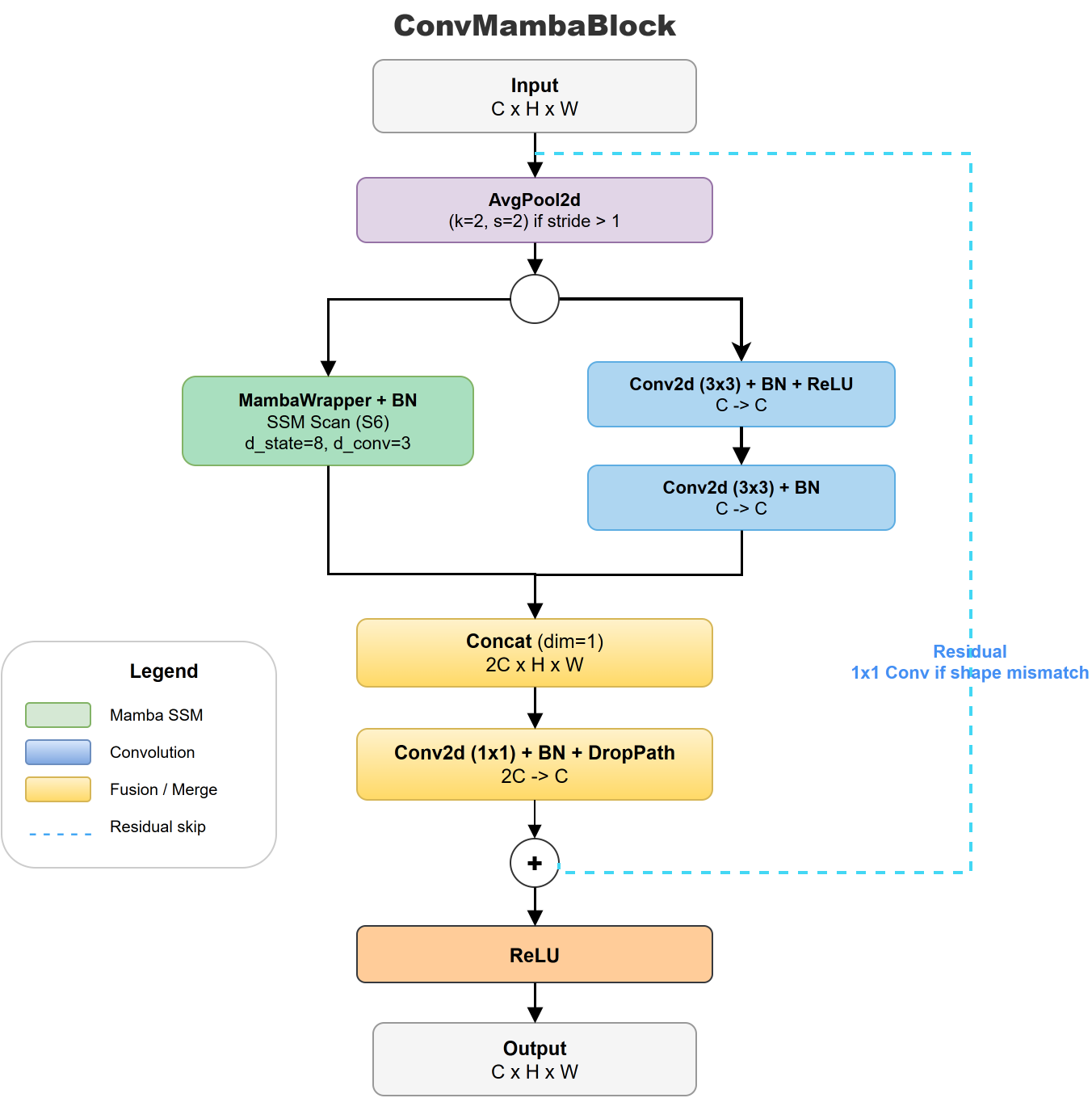}
    \caption{Architecture of the ConvMamba dual-path block. The input feature map is processed in parallel by two branches: a Mamba SSM path for global sequential dependency modeling, and a convolutional path ($3\times3 \to 3\times3$) for local spectro-temporal pattern extraction. Branch outputs are concatenated and fused via a $1\times1$ convolution. A residual connection with stochastic depth regularization links the (optionally downsampled) input to the final output.}
    \label{fig:convmamba}
\end{figure}

While the Mamba block excels at global context aggregation, it lacks the inductive bias necessary for capturing fine-grained local spectro-temporal patterns—a strength inherent to convolutions. To synergize these complementary capabilities, we propose the \textbf{ConvMamba dual-path block} (\texttt{ConvMambaBlock}), illustrated in Fig.~\ref{fig:convmamba}. 
Specifically, the block comprises two parallel branches: a Mamba branch (\texttt{MambaWrapper} + BatchNorm2d) and a convolutional branch.

Given an input $\mathbf{x} \in \mathbb{R}^{C \times H \times W}$, spatial downsampling, when required between stages ($\text{stride}>1$), is performed once via average pooling and shared by both branches---identical to the downsampling operation in Eq.~\eqref{eq:mamba_ds}:
\begin{equation}
    \mathbf{x}_s = \operatorname{AvgPool2d}_{2\times2}(\mathbf{x}),
    \label{eq:conv_downsample}
\end{equation}
where $\mathbf{x}_s = \mathbf{x}$ when $\text{stride}=1$. The two branches then compute their respective representations from this shared downsampled input as follows.

\textbf{Mamba branch:}
\begin{equation}
    \mathbf{x}_{\mathrm{m}} = \operatorname{BN}\!\bigl(\operatorname{MambaWrapper}(\mathbf{x}_s)\bigr),
    \label{eq:mamba_path}
\end{equation}

\textbf{Convolutional branch:}
\begin{equation}
    \begin{split}
        \mathbf{x}_{\mathrm{c}} &= \operatorname{BN}\!\Bigl(\operatorname{Conv}_{3\times3}\!\bigl( \\
        &\qquad \operatorname{ReLU}\!\bigl(\operatorname{BN}(\operatorname{Conv}_{3\times3}(\mathbf{x}_s))\bigr)\bigr)\Bigr).
    \end{split}
    \label{eq:conv_path}
\end{equation}

The branch outputs are concatenated along the channel dimension and fused via a $1\times1$ convolution:
\begin{equation}
    \mathbf{x}_{\mathrm{f}} = \operatorname{BN}\!\bigl(\operatorname{Conv}_{1\times1}([\mathbf{x}_{\mathrm{m}};\, \mathbf{x}_{\mathrm{c}}])\bigr),
    \label{eq:fusion}
\end{equation}
where $[\cdot;\,\cdot]$ denotes channel-wise concatenation. Finally, the output is obtained through a residual connection with stochastic depth:
\begin{equation}
    \mathbf{x}_{\mathrm{out}} = \operatorname{ReLU}\!\bigl(\operatorname{DropPath}(\mathbf{x}_{\mathrm{f}}) + \mathbf{x}_{\mathrm{skip}}\bigr),
    \label{eq:residual}
\end{equation}
where the identity shortcut follows the same projection strategy as Eq.~\eqref{eq:mamba_skip}: $\mathbf{x}_{\mathrm{skip}} = \operatorname{BN}\!\bigl(\operatorname{Conv2d}_{1\times1,\,s=2}(\mathbf{x})\bigr)$ when $\text{stride}>1$, and $\mathbf{x}_{\mathrm{skip}} = \mathbf{x}$ otherwise. Notably, the $1\times1$ fusion convolution adaptively reweights the contributions of global (Mamba) and local (convolutional) features at each spatial location, enabling data-driven integration of complementary representations.

\subsection{Multi-Scale Fusion Head}

Standard ResNet architectures perform classification using only final-stage features after global average pooling. However, successive downsampling operations discard fine-grained spectral details (e.g., pitch and timbre) encoded in early-stage representations. To mitigate this information loss, we adopt a multi-scale fusion head inspired by hierarchical feature aggregation, which integrates features from all four stages.

Let $\mathbf{F}_k \in \mathbb{R}^{C_k \times H_k \times W_k}$ denote the output feature map of stage $k$ ($k=1,\dots,4$). Each $\mathbf{F}_k$ is first spatially aligned to a uniform resolution via adaptive average pooling, then concatenated along the channel dimension:
\begin{align}
    \mathbf{F}_{\text{concat}} = \operatorname{Concat}\bigl(&\operatorname{AdaptivePool}(\mathbf{F}_1), \notag \\
    &\dots, \operatorname{AdaptivePool}(\mathbf{F}_4)\bigr).
    \label{eq:ms_concat}
\end{align}
The fused representation is subsequently compressed by a $1\times1$ convolution with BatchNorm and ReLU activation, followed by dropout regularization, global average pooling, a 128-dimensional fully connected layer, and finally the classification head.

\section{Experiments}
\label{sec:experiments}

We evaluate ResMamba in two depth configurations that differ only in the number of blocks per stage: a compact configuration with $[2,2,2,2]$ blocks and a deeper configuration with $[3,4,6,3]$ blocks. Our main experiments employ the compact configuration, instantiated with each of the three block types described in Sec.~\ref{sec:method}: Mamba-only, ConvMamba, and ConvMamba with multi-scale fusion (ConvMamba+MS). The deeper $[3,4,6,3]$ configuration, combined with the best-performing block type (ConvMamba+MS), is evaluated only in the ablation study to assess the effect of network depth (Table~\ref{tab:conv_branch_ablation}).

All models are evaluated under consistent settings with identical preprocessing and core training hyperparameters, except for dataset-specific adaptations in batch size, learning-rate schedule, and early stopping (highlighted in Table~\ref{tab:exp_settings}). Specifically, we employ five-fold stratified cross-validation on GTZAN and stratified 10-fold track-level cross-validation on FMA-Small, while FMA-Medium follows the official artist-level train/validation/test splits. These protocols are chosen to align with those of the published studies. On FMA-Medium, we use the official artist-level split released by Defferrard \textit{et al.}~\cite{fma}, the same partition on which they reported their widely cited 63\% SVM baseline. On GTZAN and FMA-Small, we adopt track-level protocols matching the reference studies~\cite{heakl2022,muniz2023}: each track is assigned intact to a single fold, so no audio content is shared between training and test sets. 

Unless otherwise noted, all experiments use full-resolution $128 \times 1292$ mel-spectrograms as input, without temporal downsampling or data augmentation. This setting preserves both the complete 128-band frequency axis and the full temporal context; its effectiveness is validated by the input-resolution ablation in Sec.~\ref{sec:ablation}, which demonstrates that accuracy improves monotonically with temporal resolution. Finally, because some published results rely on pretrained models or external data, such methods are discussed in the text (Sec.~\ref{sec:fma_comparison}) rather than used as direct baselines in the comparison tables. For every method listed in the tables, the evaluation protocol is explicitly reported.

\subsection{Dataset and Preprocessing}
\label{sec:dataset}

We evaluate all models on the GTZAN dataset~\cite{gtzan}, which comprises 1,000 audio tracks evenly distributed across ten genres: blues, classical, country, disco, hip-hop, jazz, metal, pop, reggae, and rock. Each track is stored as a 22050\,Hz mono WAV file with 16-bit resolution and a nominal duration of 30 seconds. A detailed analysis of actual track durations reveals a range from 29.93\,s to 30.65\,s, with a mean of 30.02\,s ($\sigma = 0.08$\,s). Only nine of the 999 usable tracks (0.9\%) are slightly shorter than 30\,s, 989 (99.0\%) marginally exceed this length, and one track has a duration equal to 30\,s within measurement precision.

To assess generalization beyond GTZAN, we additionally evaluate on the Free Music Archive (FMA) dataset~\cite{fma}, which offers a more realistic and challenging benchmark. Its small subset (FMA-Small) contains 8,000 tracks uniformly distributed across eight genres, while the medium subset (FMA-Medium) comprises 25,000 tracks spanning 16 genres (the 8,000 FMA-Small tracks plus 17,000 additional tracks). For FMA-Small, we employ stratified 10-fold cross-validation over track-level splits, preserving genre proportions in each fold, in line with the 10-fold, track-level evaluation commonly used in FMA-Small studies~\cite{heakl2022,muniz2023}. For FMA-Medium, since the eight FMA-Small genres are a subset of the sixteen FMA-Medium genres, we merge the two subsets by their corresponding official artist-level splits into a single 16-class protocol. All FMA tracks are provided as 30-second clips as well and undergo the identical preprocessing pipeline described below (consistent mel-spectrogram parameters), with only the classification head adjusted to match the respective number of genres.

\textbf{Audio Preprocessing.} Since all tracks in GTZAN, FMA-Small, and FMA-Medium are approximately 30 seconds long, we adopt the same input convention for all three datasets: tracks shorter than 30\,s are zero-padded at the end, while longer tracks are truncated to their first 30\,s. Mel-spectrograms are computed with $n_{\text{fft}} = 2048$, hop length $= 512$, and $n_{\text{mels}} = 128$, yielding a $128 \times 1292$ time--frequency representation for each 30-second segment. The resulting spectrograms are converted to the decibel scale and replicated across three channels to simulate RGB input. All models are trained using this identical pipeline and receive the full-resolution $128 \times 1292$ mel-spectrograms directly, without temporal downsampling. No data augmentation---neither waveform-level (e.g., pitch shifting, time stretching, or noise injection) nor spectrogram-level (e.g., SpecAugment time/frequency masking)---is applied to any input in any experiment.

\begin{figure*}[t]
\centering
\includegraphics[width=0.95\textwidth]{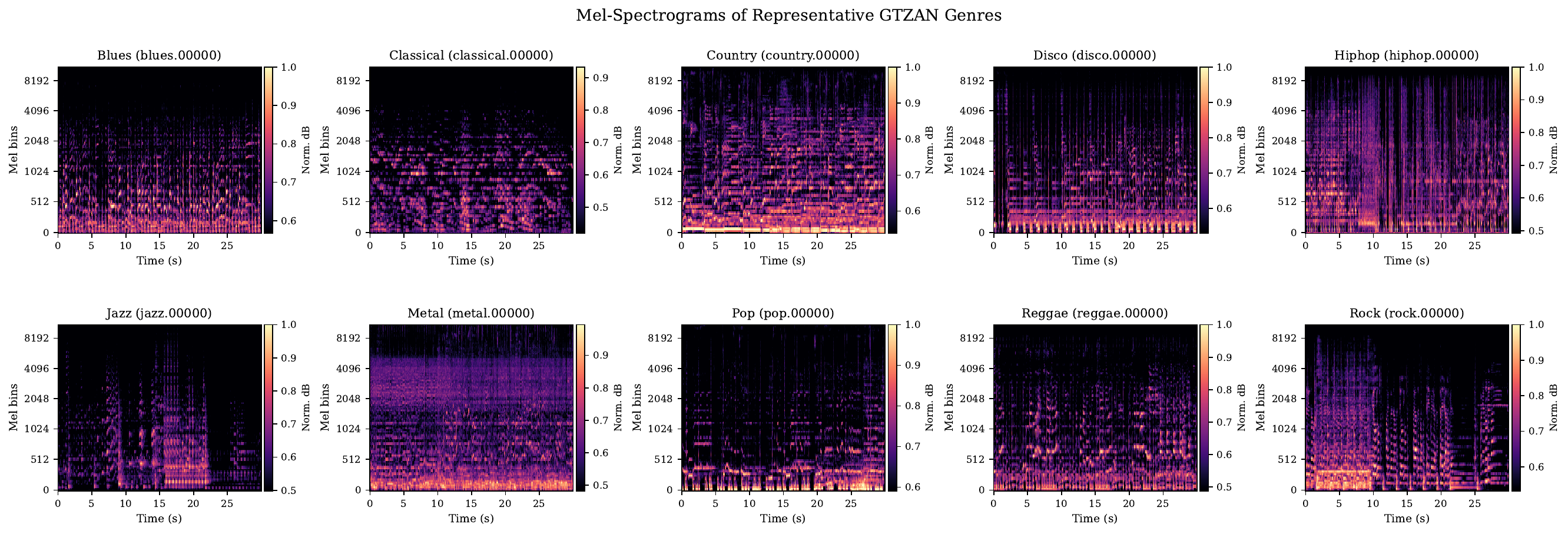}
\caption{Mel-spectrograms of the first track from each of the ten GTZAN genres: blues.00000, classical.00000, country.00000, disco.00000, hiphop.00000, jazz.00000, metal.00000, pop.00000, reggae.00000, and rock.00000. Each spectrogram is a $128 \times 1292$ decibel-scaled time--frequency representation computed with $n_{\text{fft}}=2048$, hop length $=512$, and $n_{\text{mels}}=128$. The distinct visual patterns across genres support treating genre classification as an image-based task.}
\label{fig:mel_spectrograms}
\end{figure*}

Figure~\ref{fig:mel_spectrograms} presents, for each of the ten GTZAN genres, the mel-spectrogram produced by our preprocessing pipeline for the first track of that genre in the dataset's ordering. As a compact time--frequency representation, the mel-spectrogram captures both harmonic content (vertical structure across mel bins) and temporal evolution (horizontal axis). Visually, the ten genres exhibit distinctly characteristic patterns: classical music features sparse, well-separated harmonic lines interspersed with extended silent intervals; metal and rock display dense, sustained high-energy activations across a broad frequency range; hip-hop and pop show pronounced low-frequency energy concentrated in rhythmic bursts; jazz and blues present moderate, variable textures with irregular note onsets; country and reggae demonstrate a balanced mixture of harmonic and percussive elements; and disco is dominated by a steady four-on-the-floor bass-drum pulse accompanied by lush string and brass layers. These marked visual differences suggest that music genre classification can be effectively formulated as an image recognition task, wherein deep visual backbones learn discriminative features directly from the spectrogram representation.

\textbf{Data Splitting.} For GTZAN, we employ five-fold stratified cross-validation on 999 usable tracks (one unreadable file excluded), where each fold contains approximately 800 training and 200 validation samples, with genre proportions preserved in both sets. For FMA-Small, we employ stratified 10-fold cross-validation over track-level splits: six corrupted tracks were excluded, leaving 7,994 effective tracks, and each fold uses roughly 800 tracks for evaluation and the remaining $\sim$7,200 for training, with genre proportions preserved in both sets. For FMA-Medium, we adopt the official artist-level train/validation/test splits: twenty corrupted tracks were excluded, yielding 19,904/2,504/2,572 effective train/validation/test tracks under the merged 16-class protocol. Within each dataset, all models trained in this work are evaluated on the identical effective track set under the same fold assignments, ensuring a fair comparison among them.

\subsection{Implementation Details}
\label{sec:impl}

All models are implemented in PyTorch and trained on a single NVIDIA GPU. We employ the AdamW optimizer~\cite{adamw} with an initial learning rate of $10^{-4}$ and a weight decay of $10^{-2}$, preceded by a 5-epoch linear warmup from $2 \times 10^{-5}$ to $10^{-4}$. On GTZAN, the learning rate follows a step-based schedule that multiplies it by $0.1$ at epochs 60 and 90 (half and three-quarters of the 120 training epochs), and no early stopping is applied, so each model runs for the full 120 epochs. On FMA-Small and FMA-Medium, a plateau-based schedule multiplies the learning rate by $0.5$ only when both the validation accuracy and the validation loss show no improvement for 10 consecutive epochs, down to a floor of $10^{-6}$ (1\% of the initial learning rate), and early stopping with a patience of 20 epochs is applied. Each model is trained for at most 120 epochs using the cross-entropy loss function, with a batch size of 16 on GTZAN and 32 on FMA.

For established vision backbones, we adopt the official \texttt{torchvision}~\cite{torchvision} implementations and train them from scratch without pretrained weights, modifying only the final fully connected layer for these backbones by replacing it with a randomly initialized linear classifier whose output dimension matches the number of genres in each dataset. This ensures a purely architectural comparison, largely free from confounding effects introduced by preprocessing variations. Table~\ref{tab:exp_settings} summarizes the key experimental settings for reference.

\begin{table}[t]
\centering
\caption{Summary of experimental settings.}
\label{tab:exp_settings}
\setlength{\tabcolsep}{6pt}
\small
\fitcolumn{%
\begin{tabular}{@{}lll@{}}
\toprule
\textbf{Category} & \textbf{Parameter} & \textbf{Value} \\
\midrule
\multirow{6}{*}{\textit{Audio preprocessing}} 
& Sample rate            & 22050\,Hz \\
& $n_{\text{fft}}$       & 2048 \\
& Hop length             & 512 \\
& $n_{\text{mels}}$      & 128 \\
& Segment duration       & 30\,s \\
& Mel-spectrogram size   & $128 \times 1292$ \\
\midrule
\textit{Input}    & Resolution & $128 \times 1292$ \\
\midrule
\multirow{6}{*}{\textit{Training}}
& Optimizer              & AdamW \\
& Base learning rate     & $10^{-4}$ \\
& Weight decay           & $10^{-2}$ \\
& \cellcolor{tabhl} Batch size             & \cellcolor{tabhl} 16 (GTZAN) / 32 (FMA) \\
& Epochs             & 120 \\
& Loss function          & Cross-entropy \\
\midrule
\multirow{3}{*}{\textit{LR schedule}}
& Warmup                        & Linear, 5 epochs \\
& \cellcolor{tabhl} GTZAN                         & \cellcolor{tabhl} Step: $0.1\times$ LR at epochs 60 and 90 \\
& \cellcolor{tabhl} FMA-Small, FMA-Medium         & \cellcolor{tabhl} \begin{tabular}[t]{@{}l@{}}Plateau: $0.5\times$ LR after 10 epochs \\ without improvement (val. acc. \& loss)\end{tabular} \\
\midrule
\multirow{2}{*}{\textit{Early stopping}}
& \cellcolor{tabhl} GTZAN                          & \cellcolor{tabhl} Not applied \\
& \cellcolor{tabhl} FMA-Small, FMA-Medium          & \cellcolor{tabhl} Applied (patience 20) \\
\midrule
\multirow{2}{*}{\textit{Regularization}}
& Dropout                & 0.1 \\
& Drop-path (max)        & 0.1 \\
\midrule
\multirow{3}{*}{\textit{Evaluation}}
& Cross-validation & GTZAN: track-level stratified 5-fold CV \\
& & FMA-Small: track-level stratified 10-fold CV \\
& Official split & FMA-Medium: official artist-level split \\
\midrule
\multirow{2}{*}{\textit{Hardware}}
& GPU                    & NVIDIA RTX 4090 \\
& Platform              & PyTorch 2.10.0 + CUDA 12.6 \\
\bottomrule
\addlinespace[2pt]
\multicolumn{3}{@{}l}{\footnotesize Rows with colored background indicate dataset-specific settings that differ across datasets.}
\end{tabular}%
}
\end{table}

\subsection{Ablation Studies}
\label{sec:ablation}

\subsubsection{Effect of Input Resolution}

Before evaluating the internal components of ResMamba, we first examine the impact of temporal resolution. Using ResNet-18 as a representative baseline, we compare four candidate resolutions that share the same 128-band mel-frequency axis but differ in temporal compression:

\begin{itemize}
    \item $128 \times 320$ ($\approx 4\times$ compression): aggressive downsampling approaching the $\sim 100$\,ms perceptual threshold~\cite{zwicker1999};
    \item $128 \times 640$ ($\approx 2\times$ compression): moderate temporal compression;
    \item $128 \times 960$ ($1.35\times$ compression): fine temporal granularity;
    \item $128 \times 1292$ (no compression): full-resolution input.
\end{itemize}

Downsampled resolutions are obtained via AdaptiveAvgPool2d applied to the original $128 \times 1292$ mel-spectrogram, preserving the frequency axis without interpolation. The full-resolution input is used directly without any spatial reduction.

\begin{table}[t!]
\centering
\caption{Effect of input resolution on ResNet-18 performance (5-fold average on GTZAN). \textit{Feature Map} denotes spatial dimensions before global average pooling. The full-resolution input ($128 \times 1292$) achieves the highest accuracy and is adopted for all subsequent experiments.}
\label{tab:resolution_ablation}
\setlength{\tabcolsep}{3pt}
\footnotesize
\begin{tabular}{@{}lccc@{}}
\toprule
\textbf{Resolution} & \textbf{Comp.} & \textbf{Feature Map} & \textbf{Acc (\%)} \\
\midrule
$128 \times 320$  & $4.0\times$  & $4 \times 10$  & $78.28 \pm 2.62$ \\
$128 \times 640$  & $2.0\times$  & $4 \times 20$  & $83.18 \pm 2.45$ \\
$128 \times 960$  & $1.35\times$ & $4 \times 30$  & $84.28 \pm 1.91$ \\
$128 \times 1292$ & $1.0\times$  & $4 \times 41$  & $\mathbf{85.38 \pm 1.54}$ \\
\bottomrule
\end{tabular}
\end{table}

Table~\ref{tab:resolution_ablation} presents the results. Accuracy improves monotonically with temporal resolution, indicating that every downsampling step discards some genre-discriminative temporal information. The degradation is most substantial under aggressive downsampling to $128 \times 320$ ($78.28\%$), where the effective frame interval approaches the $\sim 100$\,ms perceptual threshold~\cite{zwicker1999}, whereas the full-resolution input achieves the best accuracy ($85.38\%$). The largest gain occurs between $128 \times 320$ and $128 \times 640$ ($+4.90$ percentage points), and the temporal axis continues to carry substantial genre-discriminative information up to the full resolution, with a further improvement of $+2.20$ percentage points from $128 \times 640$ to $128 \times 1292$. This indicates that performance has not yet saturated at $\approx 23$\,ms granularity.

Based on these findings, we adopt the full-resolution $128 \times 1292$ input as the unified setting for all subsequent experiments, as it consistently yields superior accuracy.

\subsubsection{Effect of the Proposed Components}

To isolate the contribution of each proposed component, we evaluate four configurations: (i) a \textit{Mamba-only} variant, where the ConvMamba block is replaced by a plain AdaptMambaBlock (removing the convolutional branch); (ii) the ConvMamba block without multi-scale fusion; (iii) the full ConvMamba + MS model; and (iv) a deeper $[3,4,6,3]$ variant to examine the impact of network depth. All ResMamba variants share an identical ResNet-style backbone, training hyperparameters, and input resolution of $128 \times 1292$. Table~\ref{tab:conv_branch_ablation} summarizes the 5-fold cross-validation accuracy on GTZAN, with ResNet-18 serving as the baseline.

As shown in Table~\ref{tab:conv_branch_ablation}, the Mamba-only variant ($86.79\%$) already surpasses the ResNet-18 baseline ($85.38\%$) by 1.41 percentage points, demonstrating that the selective SSM's global sequential modeling is a competitive alternative to purely convolutional feature extraction. Nevertheless, this gain remains modest, because mel-spectrogram classification relies heavily on local textural patterns—such as harmonic stacks, formant transitions, and transient onsets—that align naturally with the inductive bias of convolutional receptive fields. Despite its global context capability, pure SSM-based sequence modeling lacks an explicit spatial prior for capturing local textural patterns; consequently, the locality it learns from scratch on the limited training data (GTZAN provides only 800 training samples per fold) remains limited, motivating an explicit enhancement of its local information capture capability.

The ConvMamba block effectively addresses this limitation. Its convolutional branch provides local spectral priors via two stacked $3\times3$ convolutions (with BatchNorm and ReLU), stabilizing early-layer representations, while the Mamba branch captures complementary global dependencies. A learned $1\times1$ fusion dynamically balances these pathways. Incorporating the ConvMamba block (w/o MS) yields a gain to $88.39\%$ ($+1.60$ points over Mamba-only). Adding the multi-scale fusion head further improves accuracy to $88.99\%$ ($+0.60$ points), confirming that the synergy between the convolutional and Mamba branches enhances model performance.

The additional gain from multi-scale fusion arises from the complementary nature of hierarchical features. Early-stage features retain high spatial resolution and capture fine-grained spectral details (e.g., harmonic structure, transient attacks), whereas late-stage features encode semantic genre-level patterns. The learnable $1\times1$ fusion within the MS head enables adaptive integration of information across all four backbone stages, thereby enriching the final representation beyond what single-scale features can provide.

Notably, scaling the full ConvMamba + MS model from the compact $[2,2,2,2]$ backbone ($88.99\%$, 15.1M params) to the deeper $[3,4,6,3]$ architecture ($87.49\%$, 28.0M params) leads to slight degradation rather than improvement. This corroborates our finding in Sec.~\ref{sec:baseline_comparison} that increased capacity does not guarantee better generalization on GTZAN; with limited training data, the deeper variant exacerbates overfitting instead of capturing more discriminative structure. This observation motivates our adoption of the compact ResMamba as the primary configuration.

\begin{table}[t!]
\centering
\caption{Ablation study of the proposed components on GTZAN (5-fold average, $128 \times 1292$ mel-spectrogram input). All variants share identical backbones, hyperparameters, and input resolutions. \textit{nBlocks} denotes the number of residual blocks per stage; the first four rows adopt $[2,2,2,2]$ blocks per stage, while the deeper variant uses $[3,4,6,3]$.}
\label{tab:conv_branch_ablation}
\setlength{\tabcolsep}{4pt}
\renewcommand{\arraystretch}{1.2}
\footnotesize
\resizebox{\columnwidth}{!}{%
\begin{tabular}{@{}l c c c c@{}}
\toprule
\textbf{Model}    & \textbf{Block Type}   & \textbf{nBlocks}  & \textbf{Acc. (\%)}         & \textbf{Params (M)} \\
\midrule
ResNet-18         & BasicBlock            & $[2,2,2,2]$       & $85.38 \pm 1.54$           & 11.2 \\
ResMamba          & AdaptMamba            & $[2,2,2,2]$       & $86.79 \pm 1.03$           & 2.2  \\
ResMamba          & ConvMamba             & $[2,2,2,2]$       & $88.39 \pm 2.55$           & 14.6 \\
\textbf{ResMamba} & \textbf{ConvMamba+MS} & $[2,2,2,2]$       & $\mathbf{88.99\pm1.58}$  & 15.1 \\
ResMamba          & ConvMamba+MS          & $[3,4,6,3]$       & $87.49 \pm 0.88$           & 28.0 \\
\bottomrule
\end{tabular}%
}
\end{table}

\subsection{Comparison with Established Vision Backbones}
\label{sec:baseline_comparison}

To contextualize ResMamba's design choices, we benchmark 12 established vision architectures spanning diverse design paradigms: \textbf{residual CNNs} (ResNet-18/34/50/101~\cite{resnet}), the \textbf{EfficientNet family} (B0--B3, V2-S)~\cite{efficientnet,efficientnetv2} with depthwise separable convolutions, inverted bottlenecks, and compound scaling, the \textbf{grouped-convolution network} ResNeXt-50 (32$\times$4d)~\cite{resnext}, the window-based \textbf{vision Transformer} Swin-T~\cite{swin}, and the hybrid \textbf{Mamba--Transformer} architecture MambaVision-T~\cite{mambavision}. All baselines are drawn from \texttt{torchvision}'s official model zoo except MambaVision-T, which we adopt from the official implementation released by its authors. For families with a clear capacity ladder (ResNet and EfficientNet), we include several sizes to probe whether larger models yield commensurate gains on music classification, retaining the configurations that attained the most competitive accuracy under the computational budget. Every model is trained from scratch under the identical 5-fold protocol without data augmentation and directly consumes the full-resolution $128 \times 1292$ mel-spectrogram input. In particular, the official \texttt{torchvision} Swin-T automatically zero-pads feature maps to multiples of the window size (e.g., the stage-1 feature width $1292/4 = 323$, which is not divisible by 7, is padded to $329$), so it processes the spectrogram at full resolution without resizing.

MambaVision-T is the only baseline that requires window configuration adaptation. Its default \texttt{[8, 8, 14, 7]} configuration is designed for $224 \times 224$ images: the first two stages are purely convolutional, and the last two stages use $14 \times 14$ and $7 \times 7$ windows. In our spectrogram setting, however, the stride-4 convolutional stem and stage-wise $2\times$ downsampling reduce the feature-map height of these two stages to $8$ and $4$, respectively---both below the corresponding window size---so windowing would require heavy zero-padding that leaves most windowed tokens empty. We therefore set \texttt{window\_size} $= 0$ in the last two stages: as in our model, no windowing is applied and the mixers perform global modeling over the entire feature map. The first two stages, being purely convolutional, likewise perform no windowing, making the leading \texttt{8}s in \texttt{[8, 8, 0, 0]} inert placeholders. MambaVision-T thus consumes the full-resolution $128 \times 1292$ input directly, without window-based zero-padding or resizing.

Table~\ref{tab:baseline_models} reports the 5-fold average validation accuracy and number of parameters for each model, where \textbf{ResMamba} denotes our full proposed model, i.e., the four-stage backbone with ConvMamba blocks and the multi-scale fusion (MS) head. Results confirm several observations:

\begin{itemize}
    \item \textbf{Smaller models suffice for GTZAN.} Although the ResNet variants span two block designs---the BasicBlock in ResNet-18/34 and the Bottleneck in ResNet-50/101---and parameter counts from 11.2 to 42.5 M, their accuracies (ResNet-18: 85.38\%, ResNet-34: 82.38\%, ResNet-50: 85.08\%, ResNet-101: 82.18\%) differ by at most $3.2$ points, and within each block-design family accuracy even declines with depth. Larger models thus do not necessarily help on this dataset. This justifies our choice of a compact 4-stage backbone with only two blocks per stage.
    
    \item \textbf{Efficiency-oriented designs transfer poorly to spectrograms.} All EfficientNet variants (B0--B3, V2-S), despite their compound scaling and depthwise separable convolutions, attain only 76.68--81.48\%, well below the ResNet family, suggesting that architectures optimized for natural images do not transfer directly to mel-spectrogram classification.
    
    \item \textbf{Transformers underperform without pretraining.} Swin-T, which relies on window-based self-attention and relative position biases, achieves 77.48\% despite 27.5 M parameters, the second-lowest accuracy in the table, confirming that pure attention architectures struggle to learn from the 800 training samples per fold. MambaVision-T (83.59\%) improves over Swin-T by re-introducing convolution priors, yet still lags the strongest CNNs (ResNet-18: 85.38\%, ResNet-50: 85.08\%, ResNeXt-50: 84.88\%).
    
    \item \textbf{ConvMamba's efficiency advantage.} ResMamba achieves 88.99\% with 15.1 M parameters and 15.6 G FLOPs (measured at the full $128 \times 1292$ resolution). It outperforms the strongest CNN baseline, ResNet-18 (85.38\%), by 3.61 percentage points, and the ResNeXt-50 baseline (84.88\%) by 4.11 points while using 34\% fewer parameters, delivering global context modeling at a computational cost below quadratic-complexity self-attention models (e.g., Swin-T: 19.7 G FLOPs).
\end{itemize}

\begin{table}[t!]
\centering
\caption{5-fold cross-validation results on GTZAN with $128 \times 1292$ mel-spectrogram input. Identical preprocessing and training hyperparameters are used throughout, with no data augmentation applied in any setting. Here and in the following tables, \textbf{ResMamba} denotes our full proposed model, i.e., the four-stage backbone with ConvMamba blocks and the multi-scale fusion (MS) head, with $[2,2,2,2]$ residual blocks per stage. Params are reported for the complete model (including the 10-way classification head). FLOPs are measured at the full $128 \times 1292$ resolution.}
\label{tab:baseline_models}
\setlength{\tabcolsep}{3pt}
\footnotesize
\fitcolumn{%
\begin{tabular}{@{}l c c c@{}}
\toprule
\textbf{Model} & \textbf{Acc (\%)} & \textbf{Params (M)} & \textbf{FLOPs (G)} \\
\midrule
ResNet-18                      & $85.38 \pm 1.54$ & 11.2  & 12.1 \\
ResNet-34                      & $82.38 \pm 0.95$ & 21.3  & 24.3 \\
ResNet-50                      & $85.08 \pm 1.16$ & 23.5  & 27.2 \\
ResNet-101                     & $82.18 \pm 2.71$ & 42.5  & 51.8 \\
EfficientNet-B0                & $77.98 \pm 1.39$ & 4.0  & 2.6 \\
EfficientNet-B1                & $76.68 \pm 2.98$ & 6.5  & 3.9 \\
EfficientNet-B2                & $78.38 \pm 2.16$ & 7.7  & 4.5 \\
EfficientNet-B3                & $78.48 \pm 2.78$ & 10.7  & 6.5 \\
EfficientNet~V2-S              & $81.48 \pm 1.88$ & 20.2  & 19.0 \\
ResNeXt-50 (32$\times$4d)      & $84.88 \pm 3.07$ & 23.0  & 28.2 \\
Swin-T                         & $77.48 \pm 1.75$ & 27.5 & 19.7 \\
MambaVision-T                  & $83.59 \pm 4.16$ & 31.1  & 29.6 \\
\midrule
\textbf{ResMamba}                       & $\mathbf{88.99\pm1.58}$ & 15.1 & 15.6 \\
\bottomrule
\end{tabular}%
}
\end{table}

\subsection{Comparison with State-of-the-Art Methods}
\label{sec:sota_comparison}

Table~\ref{tab:sota_comparison} compares our ResMamba (ConvMamba + Multi-Scale Fusion) against published results on the GTZAN benchmark. The cited methods span the main deep-learning families for music genre classification: attention-based vision transformers~\cite{wu2025vit}, reduced convolutional recurrent neural networks~\cite{setiorini2025}, CNN--Transformer hybrids~\cite{chen2024cnnte}, and decomposed-feature gradient boosting~\cite{gan2024vmd}.

\begin{table}[t!]
\centering
\caption{Comparison with published methods on GTZAN. Results are cited from the respective publications. The Split column states the data-splitting protocol where explicitly reported in the source paper. Other experimental settings (preprocessing, segment duration, augmentation) also vary across works.}
\label{tab:sota_comparison}
\setlength{\tabcolsep}{3pt}
\footnotesize
\fitcolumn{%
\begin{tabular}{@{}l c l@{}}
\toprule
\textbf{Method} & \textbf{Acc (\%)} & \textbf{Split} \\
\midrule
Chen~\textit{et al.}~\cite{chen2024cnnte}       & 87.41 & 10-fold, segment-level$^{a}$ \\
Gan~\textit{et al.}~\cite{gan2024vmd}           & 85.5  & 80/20 train/test$^{b}$ \\
Wu~\textit{et al.}~\cite{wu2025vit}          & 86.8  & 80/20 train/val$^{b}$ \\
Setiorini~\textit{et al.}~\cite{setiorini2025} & 88.64 & 60/20/20, segment-level$^{a}$ \\
\midrule
\textbf{ResMamba} (5-fold CV)        & $\mathbf{88.99\pm1.58}$ & 5-fold CV$^{c}$ \\
\bottomrule
\end{tabular}%
}
\par\vspace{2pt}
\begin{minipage}{\columnwidth}
\footnotesize
$^{a}$Segment-level split: short clips from the same recording are treated as independent samples, so clips of one song may appear in both training and test sets, which typically inflates accuracy relative to track-level evaluation.\\
$^{b}$Random single track-level partition without artist isolation.\\
$^{c}$Track-level stratified 5-fold cross-validation on 999 usable tracks (mean~$\pm$~std over folds).
\end{minipage}
\end{table}

ResMamba achieves the highest accuracy in Table~\ref{tab:sota_comparison} at $88.99\% \pm 1.58\%$, with a standard deviation derived from five-fold cross-validation. It surpasses the strongest published method, the reduced-CRNN model of Setiorini~\textit{et al.}~\cite{setiorini2025} (88.64\%), by 0.35 percentage points, and outperforms the attention-based method~\cite{wu2025vit} (86.8\%) by 2.19 percentage points. Against the recent baselines spanning decomposed-feature boosting~\cite{gan2024vmd} and CNN--Transformer hybrids~\cite{chen2024cnnte} (85.5--87.41\%), ResMamba gains 1.58--3.49 percentage points. 
These gains further validate the effectiveness of combining the ConvMamba dual-path design with multi-scale hierarchical feature aggregation for mel-spectrogram classification. 
It is worth noting that Yang~\textit{et al.}~\cite{yang2025informatica} (82.4\%) evaluate on GTZAN under a hierarchical 80/20 split with artist-level filtering and near-duplicate removal, which is not directly comparable to the results in the table and is therefore not tabulated here.
We note that direct numerical comparisons across published methods should be interpreted with caution due to differences in data splitting (e.g., random track-level vs.\ segment-level), segment duration, input resolution, and the presence or absence of a dedicated validation set. Our 5-fold stratified cross-validation and fixed experimental hyperparameters provide a transparent evaluation framework that facilitates meaningful comparison.

\subsection{Error Analysis}

\begin{figure}[t!]
\centering
\includegraphics[width=\columnwidth]{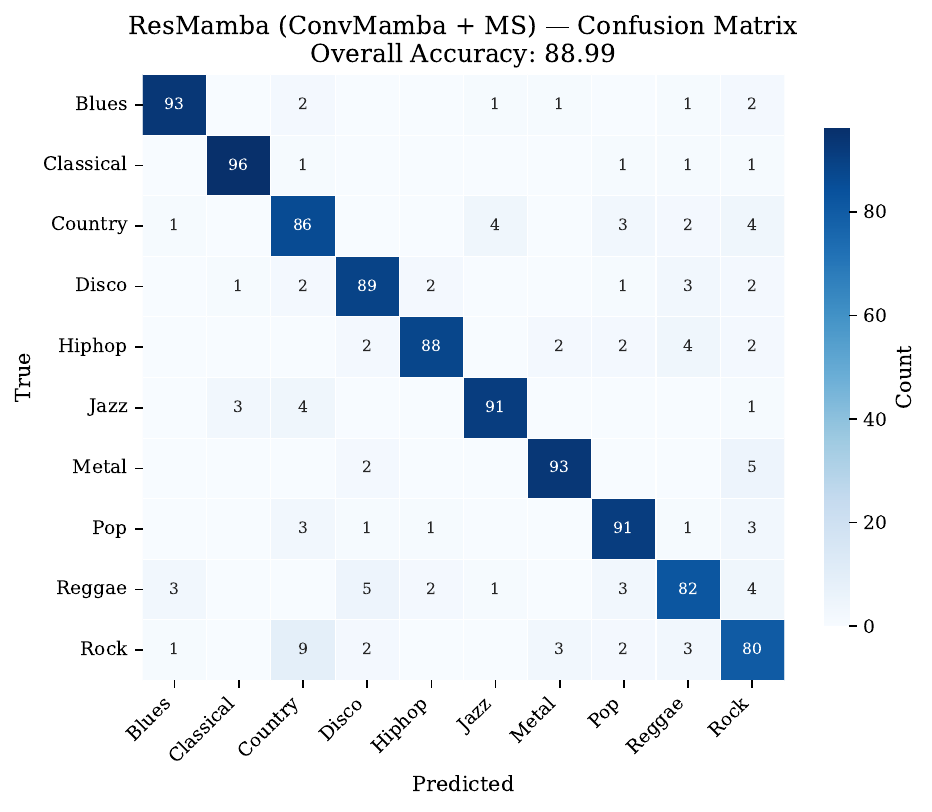}
\caption{Confusion matrix aggregated over all five cross-validation folds for ResMamba on the GTZAN dataset. Rows indicate ground-truth labels and columns indicate predictions. Diagonal entries show the number of correctly classified samples per class. The overall 5-fold accuracy is 88.99\%. Classical, Blues, and Metal achieve the highest per-class accuracy, while Rock and Reggae exhibit the most frequent inter-class confusions.}
\label{fig:confusion_matrix}
\end{figure}

To gain deeper insight into the model's behavior, we aggregate the per-sample predictions across all five cross-validation folds and construct the confusion matrix shown in Fig.~\ref{fig:confusion_matrix}.

Three genres — Classical (96/100), Blues (93/100), and Metal (93/100) — achieve the highest diagonal entries, followed closely by Jazz (91/99) and Pop (91/100), indicating that their spectro-temporal signatures are highly distinctive and well-separated from other genres. Classical music, with its sparse harmonic lines and extended silent intervals, presents a unique visual pattern that the ConvMamba backbone learns to recognize reliably. Metal and Jazz, despite both involving dense instrumentation, occupy complementary spectral regions: Metal is characterized by sustained broadband distortion, whereas Jazz exhibits transient note onsets with large dynamic range, enabling clean separation.

\setlength{\abovecaptionskip}{0pt}

Conversely, several pairs exhibit notable confusion. Rock obtains the lowest diagonal count (80/100), with its 20 errors distributed across Country (9), Metal (3), Reggae (3), Disco (2), Pop (2), and Blues (1). The most prominent confusion involves the Rock--Country pair: Rock is misclassified as Country nine times and Country as Rock four times, jointly accounting for 13 of the 34 errors committed by these two genres. This bidirectional confusion reflects the stylistic continuum between blues-influenced rock and country ballads. Rock also confuses with Metal in both directions (5 samples of Metal predicted as Rock and 3 samples of Rock predicted as Metal), mirroring the shared distorted-guitar texture of hard rock and heavy metal. Reggae (82/100) is the second most confused genre, with its 18 errors spread over Disco (5), Rock (4), Blues (3), Pop (3), Hip-hop (2), and Jazz (1). The bidirectional Reggae--Disco (8) and Reggae--Rock (7) confusions again point to the shared off-beat rhythmic patterns of these dance-oriented genres. Country (86/100) is misclassified as Rock and as Jazz equally often (4 errors each), and Jazz additionally borrows from Classical (3); these pairs share sparse arrangements and melodic contours. Hip-hop (88/100) and Disco (89/100) show mild confusions with Reggae (4 and 3, respectively), consistent with the percussive, beat-driven characteristics common to these genres. These patterns align with well-known ambiguities in the GTZAN benchmark~\cite{sturm2014} and highlight opportunities for future refinement through rhythm-informed representations or genre-specific attention mechanisms.

\setlength{\abovecaptionskip}{0.5\baselineskip}

\subsection{Comparison with Recent FMA Studies}
\label{sec:fma_comparison}

To assess the generalization of ResMamba beyond GTZAN, we additionally evaluate it on the Free Music Archive (FMA) dataset~\cite{fma} introduced in Sec.~\ref{sec:dataset}. All models are trained under the protocol described in Sec.~\ref{sec:impl} (full-resolution $128 \times 1292$ mel-spectrogram input, AdamW optimizer, batch size 32), with the classification head adjusted to the corresponding number of genres.

Table~\ref{tab:fma_comparison_small} compares ResMamba with published FMA-Small results, for which we adopt the same 10-fold cross-validation protocol. The listed baselines span a 10-fold cross-validation protocol (Heakl~\textit{et al.}~\cite{heakl2022}, 58.30\%, reporting only the highest test accuracy), nested 10-fold cross-validation (Muniz~\textit{et al.}~\cite{muniz2023}, 60.30\%), and ten random 80/20 splits (Seo~\textit{et al.}~\cite{seo2024}, 57.38\,$\pm$\,0.94\%). The highest published FMA-Small accuracy of which we are aware is 64.30\%, reported by Rai \& Bhattacharjee~\cite{rai2026} using BYOL-A pretrained embeddings with an unspecified data split; since this result relies on pretraining, it is not used as a direct baseline in the table. ResMamba, evaluated under the same track-level 10-fold cross-validation protocol without any data augmentation, attains $64.52\% \pm 1.88\%$, the highest accuracy among the from-scratch methods in Table~\ref{tab:fma_comparison_small}, exceeding the pretraining-based result of Rai \& Bhattacharjee (64.30\%) by 0.22 percentage points without any pretraining or external data. It additionally outperforms the ResNet-18 baseline ($59.95\% \pm 1.81\%$), trained from scratch under the identical protocol, by 4.57 percentage points.

Table~\ref{tab:fma_comparison_medium} compares ResMamba with published FMA-Medium results, for which we adopt the official artist-level split. Defferrard~\textit{et al.}~\cite{fma} report an accuracy of 63\% on FMA-Medium using hand-crafted features with an SVM under the official split --- the most widely cited published result on the FMA dataset. Methods that exploit pretraining or external knowledge are discussed in the text instead: Ding~\textit{et al.}~\cite{ding2024} report 68.07\% by integrating a knowledge graph built from external instrument annotations (OpenMIC-2018) with audio representations, Lin~\textit{et al.}~\cite{lin2025lca} report 69.00\% with genreMERT, a music encoder pretrained on a large corpus, Kim~\textit{et al.}~\cite{kim2018www} report an F1 score of 0.6629 on FMA-Medium relying on external artist metadata and an 85/15 split of the official training set, and Shen and Xiao~\cite{shen2026} report 78.10\% with a frozen MusicFM backbone and a linear probe. Because these methods rely on external knowledge or large-scale pretraining, they are not directly comparable with our from-scratch result (65.75\%) under the official 80/10/10 artist-level split.

\begin{table}[t!]
\centering
\caption{Comparison with published methods on the FMA-Small subset. ResMamba is evaluated under 10-fold cross-validation; evaluation protocols of the other methods are detailed in the table footnotes.}
\label{tab:fma_comparison_small}
\setlength{\tabcolsep}{6pt}
\footnotesize
\fitcolumn{%
\begin{tabular}{@{}l c l@{}}
\toprule
\textbf{Method} & \textbf{Acc. (\%)} & \textbf{Split} \\
\midrule
Heakl~\textit{et al.}~\cite{heakl2022}       & 58.30 & 10-fold CV$^{a}$ \\
Muniz~\textit{et al.}~\cite{muniz2023}       & 60.30 & Nested 10-fold CV$^{b}$ \\
Seo~\textit{et al.}~\cite{seo2024}            & 57.38$\pm$0.94 & 80/20$\times$10$^{c}$ \\
\midrule
ResNet-18                                    & 59.95$\pm$1.81 & 10-fold CV$^{d}$ \\
\textbf{ResMamba}                           & \textbf{64.52$\pm$1.88} & \textbf{10-fold CV$^{e}$} \\
\bottomrule
\end{tabular}%
}
\par\vspace{2pt}
\begin{minipage}{\columnwidth}
\footnotesize
$^{a}$10-fold CV; Heakl~\textit{et al.} report the highest test accuracy.\\
$^{b}$Nested 10-fold CV: an outer 10-fold loop evaluates generalization performance, while an inner loop performs wrapper-based feature selection within each outer training fold.\\
$^{c}$Random 80/20 split repeated 10 times.\\
$^{d}$ResNet-18 trained from scratch by us under the same protocol as ResMamba.\\
$^{e}$Track-level stratified 10-fold cross-validation on 7,994 effective tracks (mean~$\pm$~std over folds).
\end{minipage}
\end{table}

\begin{table}[t!]
\centering
\caption{Comparison with published methods on the FMA-Medium subset. Results marked with $\dagger$ are obtained by our own training runs.}
\label{tab:fma_comparison_medium}
\setlength{\tabcolsep}{6pt}
\footnotesize
\fitcolumn{%
\begin{tabular}{@{}l c c l@{}}
\toprule
\textbf{Method} & \textbf{Acc. (\%)} & \textbf{Macro F1} & \textbf{Split} \\
\midrule
Defferrard~\textit{et al.}~\cite{fma}        & 63.00       & ---          & Official artist split \\
\midrule
ResNet-18$^{\dagger}$                                & 62.99 & 0.3558 & Official artist split \\
ResNeXt-50$^{\dagger}$ (32$\times$4d)                & 61.00 & 0.3516 & Official artist split \\
\textbf{ResMamba}$^{\dagger}$                           & \textbf{65.75} & \textbf{0.4262}  & \textbf{Official artist split} \\
\bottomrule
\end{tabular}%
}
\par\vspace{2pt}
\begin{minipage}{\columnwidth}
\footnotesize
$^{\dagger}$Obtained by our own training runs: all models are trained from scratch (no pretraining, no external data, no data augmentation) on full-resolution $128 \times 1292$ mel-spectrograms, with AdamW (lr$=$$10^{-4}$). Macro F1 is the 16-class macro-averaged F1, computed by us on the corresponding test sets; published methods do not report F1 (indicated by ---).
\end{minipage}
\end{table}

The comparison confirms that FMA is substantially harder than GTZAN: the official FMA benchmark of Defferrard~\textit{et al.}~\cite{fma} attains 63\% on FMA-Medium with hand-crafted features and an SVM, well below typical GTZAN accuracies of the same class of methods. This gap reflects the larger intra-genre diversity and finer-grained genre taxonomy of FMA. Under the same official artist-level split, ResMamba reaches 65.75\%, surpassing the Defferrard baseline by 2.75 percentage points while being trained end-to-end from mel-spectrograms with no hand-crafted features, no pretraining, no external data, and no data augmentation. To place this result in context, we additionally trained ResNet-18~\cite{resnet} and ResNeXt-50~\cite{resnext} from scratch under the identical protocol (full-resolution $128 \times 1292$ mel-spectrograms, AdamW, and the same regularization); these runs are marked with $\dagger$ in Table~\ref{tab:fma_comparison_medium}.

Defferrard~\textit{et al.} also organized the official CrowdAI genre recognition challenge on FMA-Medium~\cite{defferrard2018challenge}, where models were trained on the entire FMA-Medium subset rather than on a held-out training partition, and the best submission reached a macro F1 of 0.909 on 30{,}000 held-out clips sampled from FMA-large/full. Since both the training data and the evaluation protocol differ from ours, this result is not directly comparable with the accuracies above. Odle and Lin~\cite{odle2025} reproduced the GTZAN and FMA benchmarks with a single uniform pipeline and found that, in their experiments, every architecture they tested scored below commonly reported levels, which they attribute to methodological inconsistency and data leakage in earlier studies.

The strongest published results on this benchmark are obtained under less self-contained conditions that rely on external knowledge or large-scale pretraining (see the discussion above); ResMamba trails the two closest pretraining-based results (Ding~\textit{et al.}~\cite{ding2024}, 68.07\%, and Lin~\textit{et al.}~\cite{lin2025lca}, 69.00\%) by 2.32 and 3.25 percentage points, respectively, while consuming nothing beyond the audio content of FMA-Medium itself. The higher figure of Shen and Xiao~\cite{shen2026} (78.10\%) is not directly comparable because it uses a frozen backbone with a linear probe and is therefore not trained on FMA, and the remaining pretraining-based method of Kim~\textit{et al.}~\cite{kim2018www} reports only macro F1 rather than accuracy. This suggests that pretraining the proposed ConvMamba block on a larger corpus is a promising direction for closing the remaining gap.

\section{Conclusion}
\label{sec:conclusion}

In this paper, we introduced ResMamba, an architecture that harmonizes Mamba blocks with a ResNet-style hierarchical backbone for music genre classification. We constructed a 4D-compatible Mamba block by wrapping the selective SSM into a residual module, and then designed the ConvMamba dual-path block, which combines this adapted Mamba branch with a parallel convolutional branch through a learned $1\times1$ fusion. We further employed a multi-scale fusion head that aggregates hierarchical features across all backbone stages.

Experiments on the GTZAN dataset demonstrated that the ConvMamba dual-path design with multi-scale fusion achieves a 5-fold average accuracy of $88.99\% \pm 1.58\%$. Ablation studies confirmed that each component — the convolutional branch and the multi-scale head — contributes positively to overall performance. On the larger FMA benchmarks, ResMamba attains $64.52\% \pm 1.88\%$ on FMA-Small under 10-fold cross-validation and $65.75\%$ on FMA-Medium under the official artist-level split, demonstrating consistent generalization without any pretraining or external data.

Future work may explore several directions. First, integrating the ConvMamba blocks into larger backbones (e.g., ResNet-50/101 depth) may further improve accuracy on larger datasets, where the additional capacity can be exploited more effectively than on small benchmarks such as GTZAN. Second, the window-partitioning strategy could be extended with cross-window SSM scanning (inspired by VMamba's cross-scan module) to enhance inter-window information exchange. Third, transfer learning from large-scale pretrained audio models could provide stronger initial representations. Finally, extending ResMamba to other MIR tasks such as music tagging, instrument recognition, and emotion classification would further validate the generality of the architecture.

\begin{IEEEbiography}[{\includegraphics[width=1in,height=1.25in,clip,keepaspectratio]{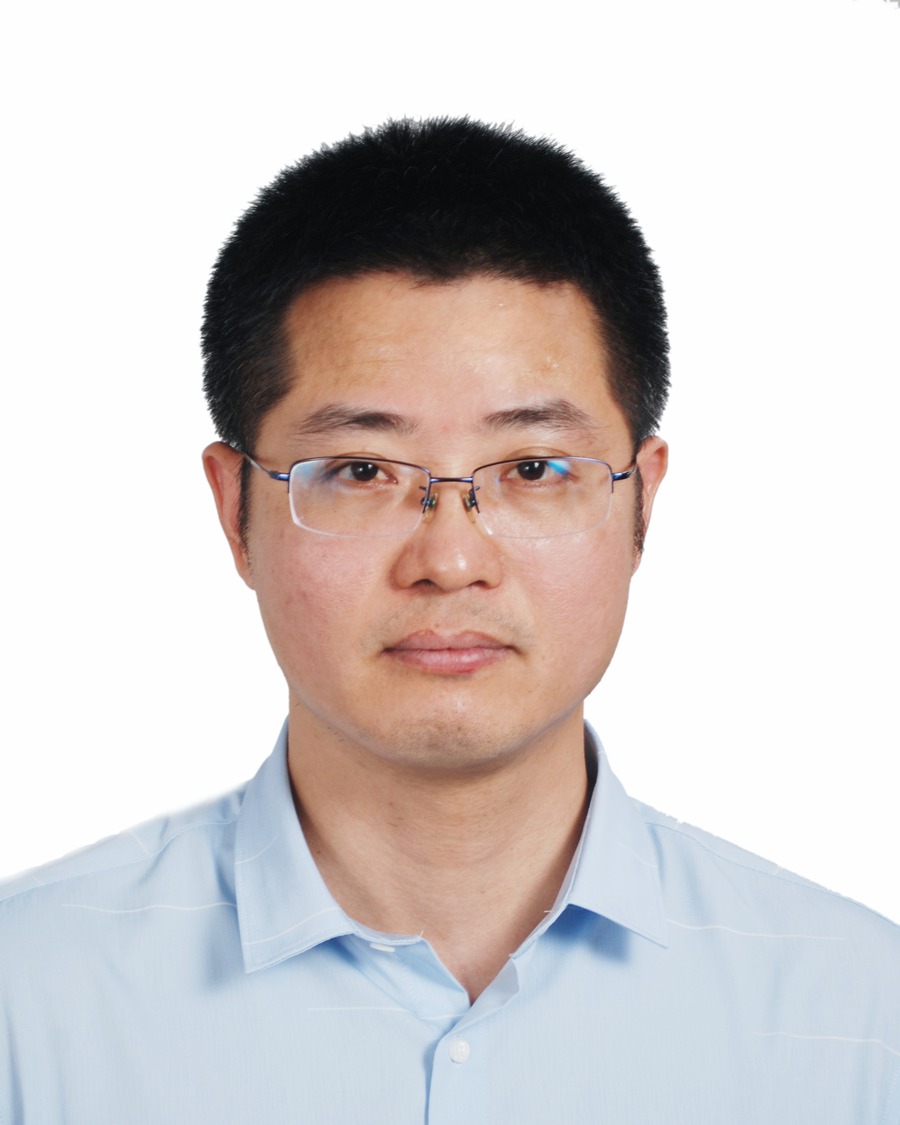}}]{Yingyu Zhang}
received the Ph.D. degree in Computer Application Technology from Huazhong University of Science and Technology, China, in 2011. He is currently an Associate Professor with the School of Artificial Intelligence, Hainan Normal University, Haikou, China. His research interests include deep learning and audio information retrieval.
\end{IEEEbiography}

\vspace{-4\baselineskip}

\begin{IEEEbiography}[{\includegraphics[width=1in,height=1.25in,clip,keepaspectratio]{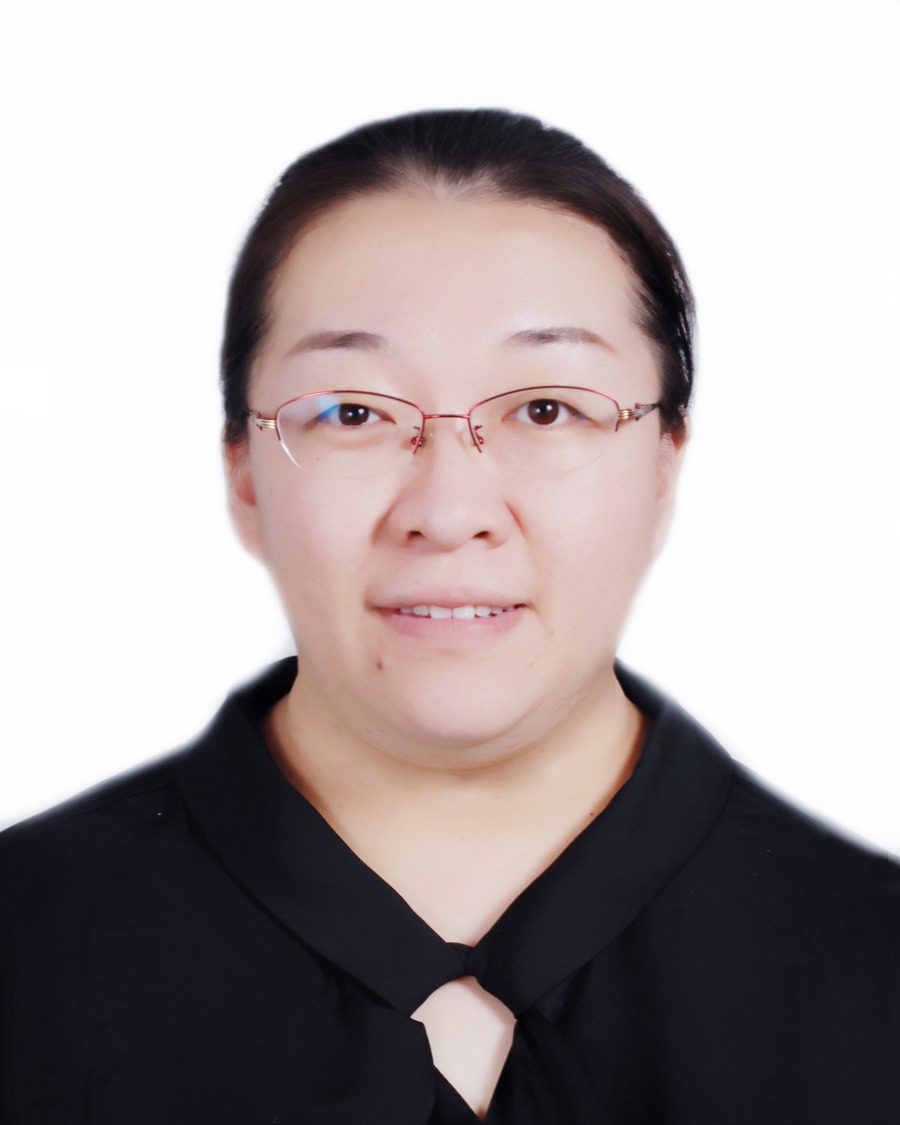}}]{Pu Qu}
received the M.S. degree in Music Education from Harbin Normal University, China. She is currently a Lecturer with the School of Teacher Education, Hainan Normal University, Haikou, China. Her research interests include music education and music aesthetics.
\end{IEEEbiography}

\EOD


\begin{thebibliography}{99}

\bibitem{gtzan}
G. Tzanetakis and P. Cook, ``Musical genre classification of audio signals,'' \emph{IEEE Trans. Speech Audio Process.}, vol. 10, no. 5, pp. 293--302, Jul. 2002. [Online]. Available: http://marsyas.info/downloads/datasets.html

\bibitem{resnet}
K. He, X. Zhang, S. Ren, and J. Sun, ``Deep residual learning for image recognition,'' in \emph{Proc. IEEE Conf. Comput. Vis. Pattern Recognit. (CVPR)}, 2016, pp. 770--778.

\bibitem{mamba}
A. Gu and T. Dao, ``Mamba: Linear-time sequence modeling with selective state spaces,'' \emph{arXiv preprint arXiv:2312.00752}, 2023.

\bibitem{s4}
A. Gu, K. Goel, and C. R\'{e}, ``Efficiently modeling long sequences with structured state spaces,'' in \emph{Proc. Int. Conf. Learn. Represent. (ICLR)}, 2022.

\bibitem{s4d}
A. Gu, K. Goel, A. Gupta, and C. R\'{e}, ``On the parameterization and initialization of diagonal state space models,'' in \emph{Proc. Adv. Neural Inf. Process. Syst. (NeurIPS)}, 2022.

\bibitem{dss}
A. Gupta, A. Gu, and J. Berant, ``Diagonal state spaces are as effective as structured state spaces,'' in \emph{Proc. Adv. Neural Inf. Process. Syst. (NeurIPS)}, 2022.

\bibitem{visionmamba}
L. Zhu, B. Liao, Q. Zhang, X. Wang, W. Liu, and X. Wang, ``Vision Mamba: Efficient visual representation learning with bidirectional state space model,'' in \emph{Proc. Int. Conf. Mach. Learn. (ICML)}, 2024.

\bibitem{mambavision}
A. Hatamizadeh and J. Kautz, ``MambaVision: A hybrid Mamba-Transformer vision backbone,'' in \emph{Proc. IEEE/CVF Conf. Comput. Vis. Pattern Recognit. (CVPR)}, 2025.

\bibitem{convmamba_hsi}
H. Zhang, H. Liu, Z. Shi, S. Mao, and N. Chen, ``ConvMamba: Combining Mamba with CNN for hyperspectral image classification,'' \emph{Neurocomputing}, 2025, Art. no. 131016.

\bibitem{medmamba}
Y. Yue and Z. Li, ``MedMamba: Vision Mamba for medical image classification,'' \emph{arXiv preprint arXiv:2403.03849}, 2024.

\bibitem{vmamba}
Y. Liu, Y. Tian, Y. Zhao, H. Yu, L. Xie, Y. Wang, Q. Ye, and Y. Liu, ``VMamba: Visual state space model,'' in \emph{Proc. Adv. Neural Inf. Process. Syst. (NeurIPS)}, 2024.

\bibitem{choi2016}
K. Choi, G. Fazekas, M. Sandler, and K. Cho, ``Convolutional recurrent neural networks for music classification,'' in \emph{Proc. IEEE Int. Conf. Acoust. Speech Signal Process. (ICASSP)}, 2017, pp. 2392--2396.

\bibitem{adamw}
I. Loshchilov and F. Hutter, ``Decoupled weight decay regularization,'' in \emph{Proc. Int. Conf. Learn. Represent. (ICLR)}, 2019.

\bibitem{hcnn}
K. Choi, G. Fazekas, and M. Sandler, ``Automatic tagging using deep convolutional neural networks,'' in \emph{Proc. Int. Soc. Music Inf. Retr. Conf. (ISMIR)}, 2016.

\bibitem{swin}
Z. Liu, Y. Lin, Y. Cao, H. Hu, Y. Wei, Z. Zhang, S. Lin, and B. Guo, ``Swin Transformer: Hierarchical vision Transformer using shifted windows,'' in \emph{Proc. IEEE/CVF Int. Conf. Comput. Vis. (ICCV)}, 2021, pp. 10012--10022.

\bibitem{crnn_music}
K. Choi, G. Fazekas, M. Sandler, and K. Cho, ``Transfer learning for music classification and regression tasks,'' in \emph{Proc. Int. Soc. Music Inf. Retr. Conf. (ISMIR)}, 2017.

\bibitem{fma}
M. Defferrard, K. Benzi, P. Vandergheynst, and X. Bresson, ``FMA: A dataset for music analysis,'' in \emph{Proc. Int. Soc. Music Inf. Retr. Conf. (ISMIR)}, 2017, pp. 316--323. [Online]. Available: https://github.com/mdeff/fma

\bibitem{defferrard2018challenge}
M. Defferrard, S. P. Mohanty, S. F. Carroll, and M. Salath\'{e}, ``Learning to recognize musical genre from audio: Challenge overview,'' in \emph{Companion Proc. Web Conf. (WWW'18 Companion)}, 2018, doi: 10.1145/3184558.3192310.

\bibitem{heakl2022}
A. Heakl, A. Abdelgawad, and V. Parque, ``A study on broadcast networks for music genre classification,'' in \emph{Proc. Int. Jt. Conf. Neural Netw. (IJCNN)}, Padua, Italy, 2022.

\bibitem{kim2018www}
J. Kim, M. Won, X. Serra, and C. C. Liem, ``Transfer learning of artist group factors to musical genre classification,'' in \emph{Proc. Web Conf. (WWW)}, 2018, pp. 1929--1934.

\bibitem{ding2024}
H. Ding, L. Zhai, C. Zhao, F. Wang, G. Wang, W. Xi, Z. Wang, and J. Zhao, ``Genre classification empowered by knowledge-embedded music representation,'' \emph{IEEE/ACM Trans. Audio, Speech, Lang. Process.}, vol.~32, pp. 2764--2781, 2024.

\bibitem{shen2024fda}
J. Shen and G. Xiao, ``Music genre classification based on functional data analysis,'' \emph{IEEE Access}, vol.~12, pp. 185483--185491, 2024.

\bibitem{chen2024cnnte}
J. Chen, X. Ma, S. Li, S. Ma, Z. Zhang, and X. Ma, ``A hybrid parallel computing architecture based on CNN and transformer for music genre classification,'' \emph{Electronics}, vol.~13, no.~16, p. 3313, 2024.

\bibitem{lin2025lca}
Y.-X. Lin, J.-C. Lin, W.-L. Wei, and J.-C. Wang, ``Learnable counterfactual attention for music classification,'' \emph{IEEE/ACM Trans. Audio, Speech, Lang. Process.}, vol.~33, pp. 570--584, 2025.

\bibitem{kim2020nc}
J. Kim, J. Urbano, C. C. S. Liem, and A. Hanjalic, ``One deep music representation to rule them all? A comparative analysis of different segmentation approaches,'' \emph{Neural Comput. Appl.}, vol.~32, no.~4, pp. 1067--1093, 2020.

\bibitem{muniz2023}
V. H. S. Muniz and J. B. O. Souza Filho, ``Robust handcrafted features for music genre classification,'' \emph{Neural Comput. Appl.}, vol.~35, pp. 9335--9348, 2023.

\bibitem{odle2025}
E. Odle and P.-C. Lin, ``Overfitting in music genre classification: Evaluating differences between the GTZAN and FMA datasets,'' in \emph{Proc. 8th Int. Conf. Artif. Intell. Cloud Comput. (AICCC)}, Tokyo, Japan, 2025, pp.~245--249.

\bibitem{shen2026}
J. Shen and G. Xiao, ``Zero-to-few shot music genre classification via prototype calibration and prompt learning in audio--language models,'' \emph{Int. J. Multimedia Inf. Retr.}, vol.~15, art. no.~24, 2026.

\bibitem{seo2024}
W. Seo, S.-H. Cho, P. Teisseyre, and J. Lee, ``A short survey and comparison of CNN-based music genre classification using multiple spectral features,'' \emph{IEEE Access}, vol.~12, pp.~245--259, 2024.

\bibitem{rai2026}
K. Rai and M. Bhattacharjee, ``Evaluating pretrained general-purpose audio representations for music genre classification,'' in \emph{Proc. Int. Conf. Pattern Recognit. Mach. Intell. (PReMI)}, 2026, pp.~840--850.

\bibitem{wu2025vit}
P. Wu, W. Gao, Y. Chen, F. Xu, Y. Ji, J. Tu, and H. Lin, ``An improved ViT model for music genre classification based on mel spectrogram,'' \emph{PLOS ONE}, 2025.

\bibitem{yang2025informatica}
Z. Yang, R. Zhang, and X. Wang, ``Music genre classification via time-frequency dual-stream neural networks and SimCLR-based self-supervised learning,'' \emph{Informatica}, vol. 49, no. 36, 2025, doi: 10.31449/inf.v49i36.10384.

\bibitem{setiorini2025}
E. Setiorini, M. Widjaja, and A. Wicaksana, ``Reduced convolutional recurrent neural network using MFCC for music genre classification on the GTZAN dataset,'' \emph{Informatica}, vol. 49, no. 17, pp. 145--154, 2025.

\bibitem{gan2024vmd}
R. Gan, T. Huang, J. Shao, and F. Wang, ``Music genre classification based on VMD-IWOA-XGBOOST,'' \emph{Mathematics}, vol. 12, no. 10, p. 1549, 2024.

\bibitem{sturm2014}
B. L. Sturm, ``The state of the art ten years after a state of the art: Future research in music information retrieval,'' \emph{J. New Music Res.}, vol. 43, no. 2, pp. 147--172, 2014.

\bibitem{torchvision}
S. Marcel and Y. Rodriguez, ``Torchvision the machine-vision package of torch,'' in \emph{Proc. ACM Int. Conf. Multimedia (MM)}, 2010, pp. 1485--1488.

\bibitem{vgg}
K. Simonyan and A. Zisserman, ``Very deep convolutional networks for large-scale image recognition,'' in \emph{Proc. Int. Conf. Learn. Represent. (ICLR)}, 2015.

\bibitem{densenet}
G. Huang, Z. Liu, L. van der Maaten, and K. Q. Weinberger, ``Densely connected convolutional networks,'' in \emph{Proc. IEEE Conf. Comput. Vis. Pattern Recognit. (CVPR)}, 2017, pp. 4700--4708.

\bibitem{mobilenetv2}
M. Sandler, A. Howard, M. Zhu, A. Zhmoginov, and L.-C. Chen, ``MobileNetV2: Inverted residuals and linear bottlenecks,'' in \emph{Proc. IEEE Conf. Comput. Vis. Pattern Recognit. (CVPR)}, 2018, pp. 4510--4520.

\bibitem{mobilenetv3}
A. Howard, M. Sandler, G. Chu, L.-C. Chen, B. Chen, M. Tan, W. Wang, Y. Zhu, R. Pang, V. Vasudevan, Q. V. Le, and H. Adam, ``Searching for MobileNetV3,'' in \emph{Proc. IEEE/CVF Int. Conf. Comput. Vis. (ICCV)}, 2019, pp. 1314--1324.

\bibitem{shufflenetv2}
N. Ma, X. Zhang, H.-T. Zheng, and J. Sun, ``ShuffleNet V2: Practical guidelines for efficient CNN architecture design,'' in \emph{Proc. Eur. Conf. Comput. Vis. (ECCV)}, 2018, pp. 116--131.

\bibitem{efficientnet}
M. Tan and Q. V. Le, ``EfficientNet: Rethinking model scaling for convolutional neural networks,'' in \emph{Proc. Int. Conf. Mach. Learn. (ICML)}, 2019, pp. 6105--6114.

\bibitem{efficientnetv2}
M. Tan and Q. V. Le, ``EfficientNetV2: Smaller models and faster training,'' in \emph{Proc. Int. Conf. Mach. Learn. (ICML)}, 2021, pp. 10096--10106.

\bibitem{resnext}
S. Xie, R. Girshick, P. Doll{\'a}r, Z. Tu, and K. He, ``Aggregated residual transformations for deep neural networks,'' in \emph{Proc. IEEE Conf. Comput. Vis. Pattern Recognit. (CVPR)}, 2017, pp. 1492--1500.

\bibitem{zwicker1999}
E. Zwicker and H. Fastl, \emph{Psychoacoustics: Facts and Models}, 2nd ed. Berlin, Germany: Springer, 1999.

\end{thebibliography}
\end{document}